\documentclass[aps,pra,twocolumn,superscriptaddress,a4paper,longbibliography,english]{revtex4-1}

\usepackage{times}
\usepackage{color}
\usepackage[colorlinks=true,urlcolor=blue,citecolor=blue]{hyperref}
\usepackage{url}
\usepackage{breakurl}
\usepackage{soul}
\usepackage{graphicx}
\usepackage{amsmath}
\usepackage{subfigure}
\usepackage{xcolor}
\usepackage{amssymb}
\usepackage{latexsym}
\usepackage{hyperref}% add hypertext capabilities
\DeclareGraphicsExtensions{.png,.eps}
\usepackage{float}
\usepackage{xcolor}
\usepackage[urlcolor=blue]{hyperref}      % links to citations
\hypersetup{
    colorlinks = true,                    % text and not border
    citecolor = {blue},
    linkcolor = {purple},
           }
\begin{document}	
	
\title{Encoding of Matrix Product States into Quantum Circuits of One- and Two-Qubit Gates} 

\author{Shi-Ju Ran}\email[Email: ]{sjran@cnu.edu.cn}
\affiliation{Department of Physics, Capital Normal University, Beijing 100048, China}
\date{\today}

\begin{abstract}
	The matrix product state (MPS) belongs to the most important models in, for example, quantum information sciences and condensed matter physics. However, realizing an $N$-qubit MPS with large $N$ and large entanglement on a quantum platform is extremely challenging, since it requires high-level qudits or $n$-qubit gates with $n \gg 2$ to carry or produce the entanglement. In this work, an efficient method that accurately encodes a given MPS into a quantum circuit with only one- and two-qubit gates is proposed. Essentially different from the existing compiling methods, our idea is to construct the unitary matrix product operators that optimally disentangle the MPS to a product state. These matrix product operators form the quantum circuit that evolves a product state to the targeted MPS with a high fidelity. Our benchmark on the ground-state MPS's of the strongly-correlated spin models show that the constructed quantum circuits can simulate the MPS's with much fewer qubits than the sizes of the MPS's themselves. This method paves a feasible and efficient path to realizing useful and/or exotic quantum states and MPS-based models as quantum circuits on the near-term quantum platforms.
\end{abstract}

\maketitle

\section{Introduction}

Matrix product state (MPS) is one of most successful mathematic tools in the contemporary physics. In condensed matter physics, MPS is the state ansatz behind the famous density matrix renormalization group (DMRG) algorithm \cite{W92DMRG, W93DMRG, DMNS98MPS} and many of its variants \cite{MC96FTDMRG, WX97TMRG, BSZ03exciteMPS, CW09EntPurt, HPWC+12MPSexcitations}. MPS can efficiently describe the ground states and (purified) thermal states of one-dimensional (1D) gapped systems \cite{V04TEBD, VGC04MPDO, VC06MPSFaithfully, V07iTEBD, LRGZXY+11LTRG}. It has also been widely and successfully applied to other areas including statistic physics \cite{JCJ10MPSstat}, non-equilibrium quantum physics \cite{VGC04MPDO, PZ09MPSsteady, WMIS14MPSDMFT, SC16MPSopen, JMC18MPSopen}, field theories \cite{CS10critical, VC10cMPS, HCOV13cMPS, MHO13MPSfield, SRHE14fieldcMPS, RGV15cMPS}, machine learning \cite{SS16TNML, HWFWZ17MPSML, CCXWX18MPSbolzmann, HPWS18TNQML, SPLRS19GTNC}, and so on.

In particular, MPS is an important model in quantum information and computation (see, e.g., \cite{VPC04DMRGQinfo, GESP07MPSQC, VWC09QCP, DHH19MPSshor}). It can represent a large class of states, including GHZ \cite{BS18MPSQC} and AKLT states \cite{AKLT87AKLTState, PVWC07MPSRev}, which can be used to implement non-trivial quantum computational tasks \cite{PhysRevLett.92.087201, GE07QCPEPS, WAR11AKLTQC, ESBD12AKLTQC}. However, realizing MPS on quantum hardwares is strictly limited. This is partially due to the fact that current techniques only permit short coherent time and small numbers of computing qubits. Solid progresses are reported in this direction recently, for instance, the realization of the GHZ state up to twenty qubits in a (relatively) long coherent time \cite{SXLZ+19GHZ20, OLKS+19CatExp}.

Moreover, MPS is hindered by another essential difficulty. There are two kinds of degrees of freedom in MPS, which are physical degrees of freedom representing the Hilbert space in which the physical model is defined, and the virtual degrees that carry the entanglement of the MPS. In general, the dimension of the virtual degrees of freedom (denoted as $\chi$) is much larger than the physical dimension (denoted as $d$). To realize an MPS in a quantum platform, one intuitively needs to realize $\chi$-level qudits as the virtual degrees of freedom \cite{PhysRevLett.95.110503}. This becomes almost impossible considering we usually take $\chi \sim O(10^2)$ or even larger. 

One way to get around the $\chi$-level qudits is to introduce multiple-qubit gates (see, e.g., \cite{cramer2010efficient, HPWS18TNQML}), where the $\chi$-level qudits in the circuits of the MPS's are equivalently replaced by several two-level qubits. Since such a scheme contains multiple-qubit gates, one must further compile these gates to one- and two-qubit gates to implement on the realistic quantum hardware \cite{BBCD+95universalGate, CFM17Qcompile}. The MPS should be in a form similar to, for instance, the state ansatz for the variational quantum eigensolvers \cite{LZWW19QMPS}. However, compiling an MPS of large $\chi$ is extremely inefficient, since the depth of the circuit generally scales polynomially with $\chi$ \cite{MV06Qgates}. Therefore, efficient encoding algorithms for MPS's are strongly desired.

\begin{figure}[tbp]
	\centering
	\includegraphics[angle=0,width=1\linewidth]{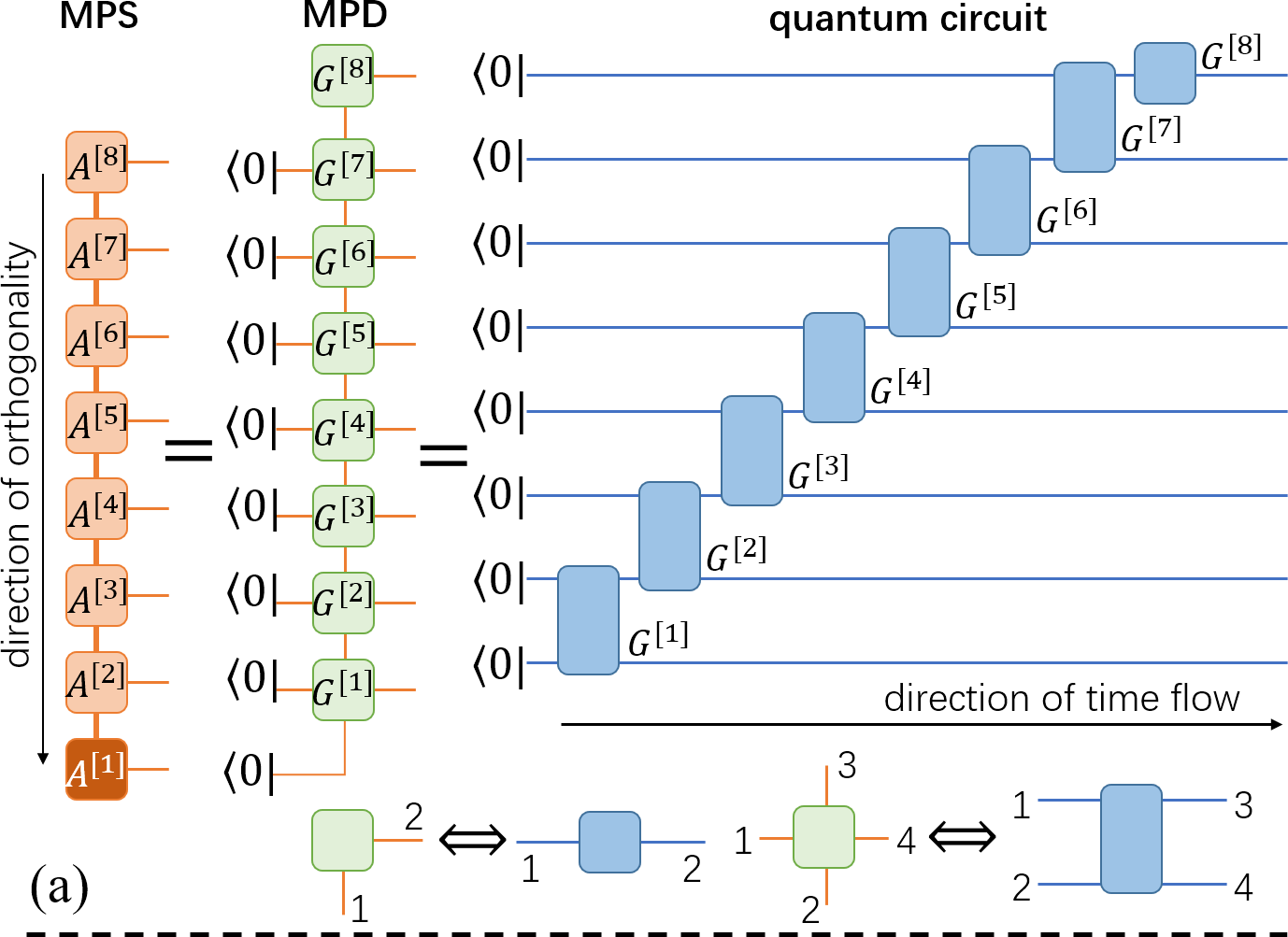}
	\includegraphics[angle=0,width=1\linewidth]{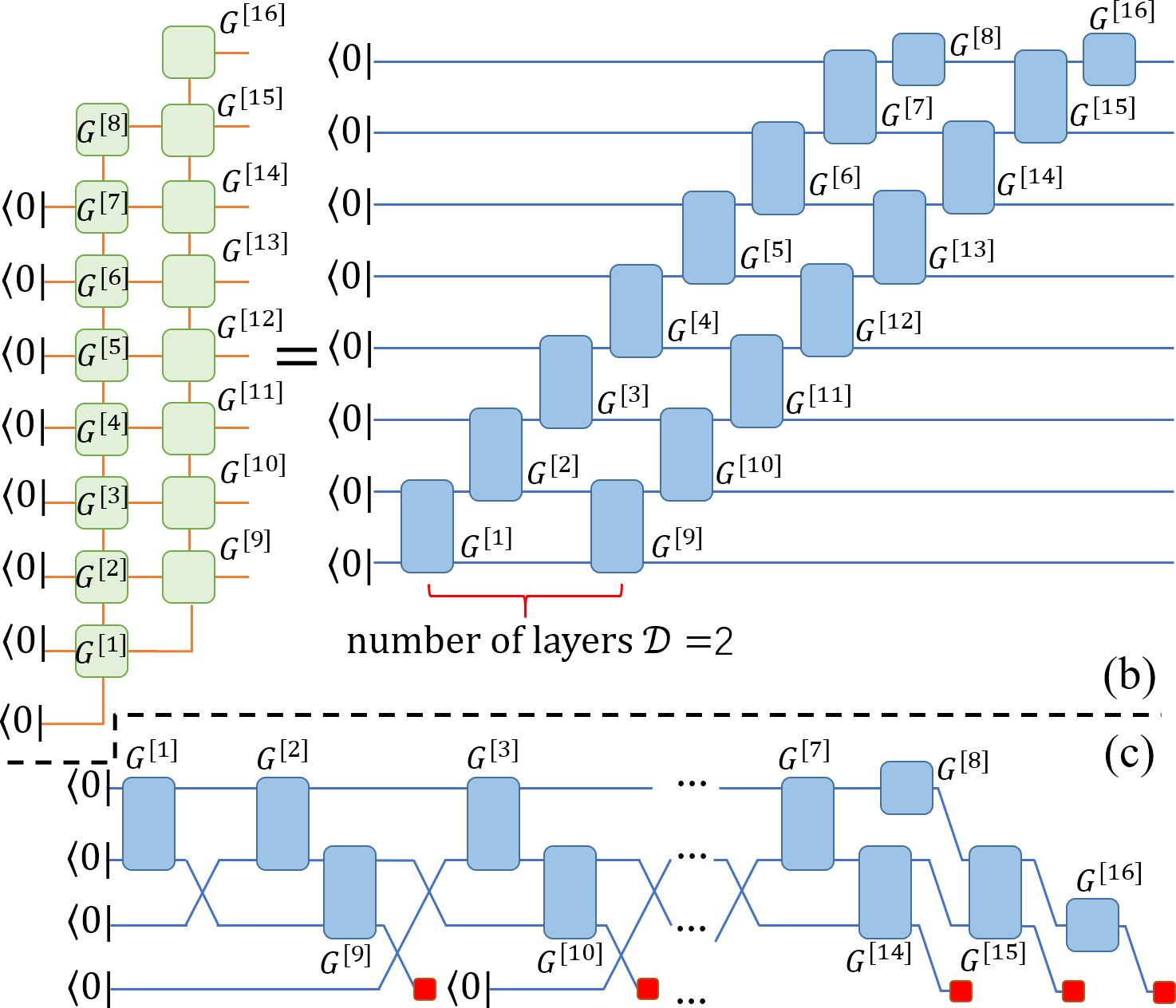}
	\caption{(Color online) (a) The diagram of $|\psi\rangle = \hat{U}^{\dagger} |0\rangle$, with $|\psi\rangle$ and $\hat{U}$ the MPS and MPD, respectively. The single-layer quantum circuit \cite{PhysRevLett.95.110503} is also illustrated. The correspondence of the indexes of the tensors in the MPD and the gates in the circuit is shown at the bottom. (b) illustrates the deep quantum circuit consisting of $\mathcal{D}=2$ layers of matrix product disentanglers. Note we take a small number of sites ($N=8$) here only for illustration. In the actual simulations, we take $N = 24 \sim 150$. (c) shows the qubit-efficient scheme \cite{HPWS18TNQML} of the deep circuit. We use the red squares to mark the degrees of freedom corresponding to the physical indexes of the MPS. The crosses of lines are swap gates.}
	\label{fig-mpd1}
\end{figure}

In this work, we propose an algorithm that efficiently and accurately encodes a given MPS with $d=2$ and $\chi \gg d$ into a quantum circuit consisting of only one- and two-qubit gates. The idea is to construct the unitary matrix product operators \cite{CPSV17UMPO, SBS+18UMPO}, dubbed as matrix product disentanglers (MPD's), that disentangle the targeted MPS [Figs. \ref{fig-mpd1} (a) and (b)]. These MPD's form a multi-layer quantum circuit, which evolves a product state into the MPS with a high fidelity. 

We testify our encoding algorithm on the MPS's that approximate the ground states of the 1D strongly-correlated spin systems. Since these MPS's possess large entanglement, it is obviously difficult to realize them on quantum circuits by the existing methods. We show that high fidelity between the MPS's and the evolved states by quantum circuits can be reached with only $O(10)$ layers of MPD's. By incorporating with the qubit-efficient scheme \cite{HPWS18TNQML}, our method efficiently encodes the MPS's into a quantum circuit of less than 10 qubits [Fig. \ref{fig-mpd1} (c)], which is much less than the size of the MPSs' themselves.

\section{Matrix product state and orthogonal form: preliminaries} 

An MPS [Fig. \ref{fig-mpd1} (a)] consisting of $N$ sites (qubits or qudits) can be written as
\begin{equation}\label{eq-MPS}
|\Psi \rangle = \sum_{a_1 \cdots a_{N-1}} \sum_{s_1 \cdots s_N} A^{[1]}_{s_1, a_1} A^{[2]}_{s_2, a_1 a_{2}} \cdots A^{[N]}_{s_N, a_{N-1}} \prod_{n=1}^N|s_n\rangle,
\end{equation} 
with the physical indexes $\{s_n = 0, \cdots, d-1\}$ and virtual indexes $\{a_n = 0, \cdots, \chi-1\}$ which label the physical and virtual degrees of freedom, respectively. The virtual dimensions are normally bounded as $\dim(a_n) \leq \chi$ to control the computational cost (see, e.g., \cite{W92DMRG, W93DMRG, V07iTEBD}), with $\chi$ called the dimension cut-off. 

Generally, one takes $\chi \gg d$ to sufficiently capture the entanglement. Take the MPS representing a one-dimensional (1D) critical state or conformal field theory as an example. One should take $\chi \sim N^{\alpha}$, with $\alpha (\simeq 1)$ determined by the scaling laws of the correlation length and entanglement entropy \cite{VLRK03CritEnt, TOIL08EntScaling}.

An MPS can be transformed into different orthogonal forms. Let us focus on the left-orthogonal form that will be used later here. $A^{[1]}$ satisfies the normalization condition, and the rest satisfy the left-orthogonal conditions, i.e., 
\begin{eqnarray}
&&\sum_{s_1a_1} A^{[1]}_{s_1, a_1} A^{[1]\ast}_{s_1, a_1} = 1,  \label{eq-cond1} \\
&&\sum_{s_na_{n}} A^{[n]}_{s_n, a_{n-1} a_{n}} A^{[n]\ast}_{s_n, a'_{n-1} a_{n}}  = I_{a_{n-1} a'_{n-1}},  \label{eq-cond2} \\
&&\sum_{s_N} A^{[N]}_{s_N, a_{N-1}} A^{[N]\ast}_{s_N, a'_{N-1}} = I_{a_{N-1} a'_{N-1}}, \label{eq-cond3}
\end{eqnarray}
with $1<n<N$ in Eq. (\ref{eq-cond2}) and $I$ the identity. The orthogonal conditions in fact determine the directions of the renormalization-group flows in the Hilbert space \cite{S11DMRGRev}. Any MPS can be transformed into the left-orthogonal form by gauge transformation.

%\begin{figure}[tbp]
%	\centering
%	\includegraphics[angle=0,width=0.7\linewidth]{mps.png}
%	\caption{(Color online) (a) The illustration of the left-orthogonal matrix product state of eight particles, where the orthogonal center is set on the first tensor; (b) the normalization condition of the tensor at the orthogonal center; (c) the orthogonal conditions of the rest of the tensors.}
%	\label{fig-MPS}
%\end{figure}

\textit{Encoding matrix product state into single-layer quantum circuit.--- } Before introducing our deep quantum circuit scheme, let us first explain the single-layer scheme proposed in Ref. \cite{PhysRevLett.95.110503} in the language of MPS and MPD. We temporarily assume $d=\chi=2$ (the reason will be explained later), and show how to exactly encode such a  left-orthogonal MPS into a quantum circuit consisting of only one- and two-body gates on the $d$-level qudits. Note a $d$-level qudit is a qubit. The given arguments can be readily generalized to $d>2$.

The MPD $\hat{U}$ that disentangles the MPS $|\psi\rangle$ into a product state is defined as
\begin{equation}\label{eq-MPD}
\hat{U} |\psi\rangle = \prod_{\otimes n=1}^N |0\rangle_n \overset{\text{def}}{=} |0\rangle.
\end{equation}
To explain how to obtain the MPD from a given MPS, we take an MPS formed by $N=8$ tensors as an example [Fig. \ref{fig-mpd1} (a)]. For the numerical simulations given later in this work, we take $N=24 \sim 150$. For the last tensor ($N=8$), we have 
\begin{equation}\label{eq-G8}
G^{[8]} = A^{[8]},
\end{equation}
which is just a one-qubit unitary gate [satisfying Eq. (\ref{eq-cond2})].

% $\hat{U}$ is unitary, i.e., $\hat{U} \hat{U}^{\dagger} = \hat{U}^{\dagger} \hat{U} = I$. Then from Eq. (\ref{eq-MPD}), one readily has $|\psi\rangle = \hat{U}^{\dagger} |0\rangle$, meaning $U^{\dagger}$ evolves the product state into the MPS.

%\begin{equation}\label{eq-MPD1}
%|\psi\rangle = \hat{U}^{\dagger} \prod_{\otimes n=1}^N |0\rangle_n.
%\end{equation}

For $1<n<N$, the tensor in $\hat{U}$ is $(d\times d\times d\times d)$, denoted as $G^{[n]}_{ijkl}$. The component of $G^{[n]}$ with $i=0$ is given by $A^{[n]}$ in the MPS, i.e.,
\begin{equation}\label{eq-Gn1}
G^{[n]}_{0jkl} = A^{[n]}_{jkl}.
\end{equation}
The components for $i=1, \cdots, d-1$ are obtained in the kernel. According to Eq. (\ref{eq-cond2}), one considers $A^{[n]}_{jkl}$ with $j=0, \cdots, d-1$ as the orthonormal vectors in the $d^2$-dimensional vector space. Then choose ($d^2-d$) orthonormal vectors in the kernel of $A^{[n]}$ as $G^{[n]}_{ijkl}$ ($i=1, \cdots, d-1$, $j\neq 0$). Together with Eq. (\ref{eq-Gn1}), we have
\begin{eqnarray}
\sum_{kl} G^{[n]}_{i'j'kl} G^{[n]\ast}_{ijkl} = I_{i'i} I_{j'j}. \label{eq-Gn2}
\end{eqnarray}
Eq. (\ref{eq-Gn2}) gives the orthonormal conditions, and means that $G^{[n]}$ is a two-qubit unitary gate.

% $G^{[n]}$ can always be defined in this way since $\dim(i) \dim(j) = \dim(k) \dim(l) = d^2$. These $d^2$ vectors just form the complete orthonormal basis in the $d^2$-dimensional vector space.

% $G^{[n]}$ can always be defined in this way since $\dim(i) \dim(j) = \dim(k) \dim(l) = d^2$. Specifically for $d=2$, $G^{[n]}_{10kl}$ and $G^{[n]}_{11kl} $ are the two normalized vectors that are orthogonal to $G^{[n]}_{00kl}$ and $G^{[n]}_{01kl}$. These four vectors just form the complete orthonormal basis of the four-dimensional vector space. Eq. (\ref{eq-Gn2}) gives the orthonormal conditions, and means that $G^{[n]}$ is a two-qubit unitary gate.

For $n=1$, the tensor is also forth-order, denoted as $G^{[1]}_{ijkl}$. The component for $i=j=0$ is given as
\begin{equation}\label{eq-G11}
G^{[1]}_{00kl} = A^{[1]}_{kl}.
\end{equation}
Again, $G^{[1]}_{00kl}$ is a $d^2$-dimensional normalized vector [see Eq. (\ref{eq-cond1})]. The rest ($d^2-1$) components of $G^{[1]}_{ijkl}$ (with $i \neq0$ or $j\neq 0$) are the orthonormal vectors in the kernel of $A^{[1]}$. The orthonormal conditions are the same as Eq. (\ref{eq-Gn2}), which means that $G^{[1]}$ is also a two-qubit unitary gate. In short, the MPD $\hat{U}$ can be obtained with Eqs. (\ref{eq-G8})-(\ref{eq-G11}) from a given MPS $|\psi \rangle$ . Then by definition [Eqs. (\ref{eq-G8}), (\ref{eq-Gn1}), and (\ref{eq-G11})], we have $|\psi\rangle = \hat{U}^{\dagger} |0\rangle$. Meanwhile, $\hat{U}$ is unitary \cite{note1}, and we immediately have $\hat{U} |\psi\rangle =  |0\rangle$. It means that $\hat{U}$ disentangles the MPS into a product state. Fig. \ref{fig-mpd1} (a) shows the circuit of $\hat{U}$. We can see $\dim(i) = \dim(k)$ and $\dim(j) = \dim(l)$ are required since $i$ and $k$, $j$ and $l$ represent the same qubit, respectively. This means $\chi=d$ in the MPS.

% The quantum circuit of the MPD is shown in Fig. \ref{fig-mpd1} (a). The correspondence of the indexes between the tensors of the MPD and the gates in the circuit are given at the bottom of (a). 

\begin{figure}[tbp]
	\centering
	\includegraphics[angle=0,width=0.85\linewidth]{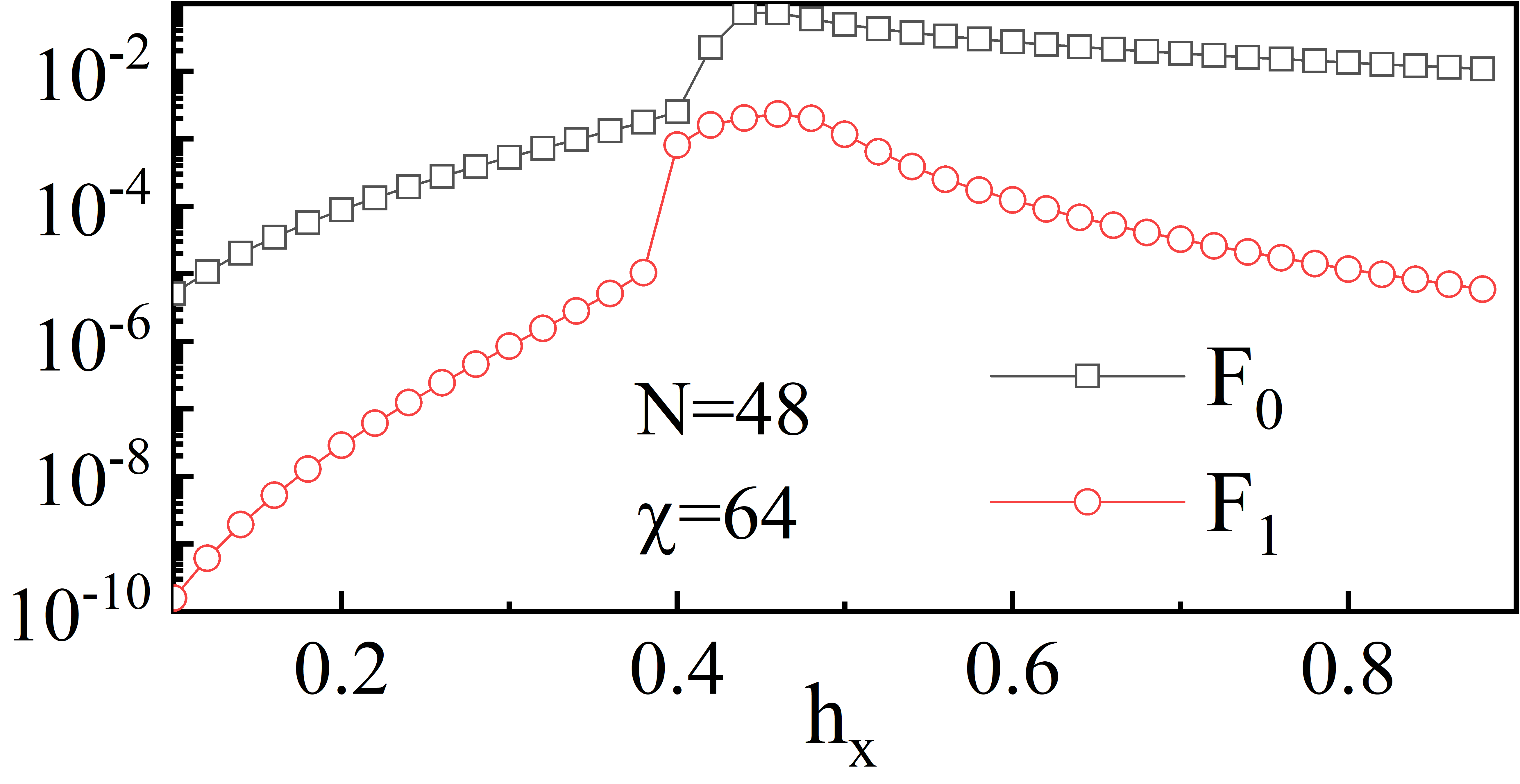}
	\caption{(Color online) Semi-log plot of the NLF's $F_0$ [Eq. (\ref{eq-fid0})] and $F_1$ [Eq. (\ref{eq-fid1})] of the ground state of transverse Ising model with different magnetic field $h_x$. $F_0$ characterize the minimal distance between $| \psi \rangle$ and a separable state. $F_1$ characterizes how accurately the quantum circuit evolves the product state $|0\rangle$ to the targeted MPS.}
	\label{fig-fid0}
\end{figure}

Different from Ref. \cite{PhysRevLett.95.110503}, we now consider $\chi > d$ and define the MPD as following. First, find the optimal MPS $|\tilde{\psi} \rangle$ with $\chi=d$ that maximizes the fidelity $|\langle \psi | \tilde{\psi} \rangle|$. This can be done by reducing the virtual dimensions of $|\psi \rangle$ to $d$ with the standard truncation algorithm of MPS (see \cite{PVWC07MPSRev} for example). Then, $\hat{U}$ can be obtained from $|\tilde{\psi} \rangle$ following the standard procedure. In this case, $\hat{U}$ cannot disentangle $|\psi \rangle$ into a product state, but will largely reduce its entanglement. To testify this statement, we consider the ground state of the 1D transverse Ising model with the Hamiltonian $\hat{H} = \sum_{n=1}^{N-1} \hat{S}^z_n \hat{S}^z_{n+1} - h_x \sum_{n=1}^N \hat{S}^x_n$. The MPS is obtained by the DMRG algorithm \cite{W92DMRG, W93DMRG} with $N=48$ and $\chi=64 \gg d$ (note $d=2$ for spin-$1/2$ models). We calculate the negative-logarithmic fidelities (NLF) per site (see Fig. \ref{fig-fid0})
\begin{eqnarray}
&&F_0 = -\frac{\ln |\langle \psi | \tilde{\psi}_{\chi=1} \rangle|}{N}, \label{eq-fid0} \\ 
&&F_1 = -\frac{\ln | \langle \psi | \hat{U}^{\dagger}  |0 \rangle|}{N}. \label{eq-fid1}
\end{eqnarray}
In $F_0$, the state $| \tilde{\psi}_{\chi=1} \rangle$ is an MPS with $\chi=1$ (a separable state) that is optimally truncated from $| \psi \rangle$. $F_0$ gives in fact the global entanglement of the MPS, which characterizes the minimal distance between $| \psi \rangle$ and a separable state \cite{WG03GE}. $F_1$ characterizes the distance between $| \psi \rangle$ and $\hat{U}^{\dagger} |0 \rangle$, i.e., how accurately $\hat{U}^{\dagger}$ evolves $|0\rangle$ to $|\psi \rangle$.

For $h_x \ll h_c$ and $h_x \gg h_c$ with $h_c=0.5$ the critical point (strictly speaking, $h_c=0.5$ is critical for $N\to \infty$ since criticality is defined in the thermodynamic limit), $|\psi \rangle$ is in the N\'eel phase and polarized phase, respectively. The ground-state entanglement in these phases is relatively small. However, $F_0$ is still non-zero in these regions due to the quantum fluctuations, which requires $\chi>1$. As expected, a peak of $F_0$ appears near the critical point, where $|\psi \rangle$ is quite far away from a separable state.

Disentangled by $\hat{U}$, the state becomes much closer to the product state $|0\rangle$, where $F_1$ is about $O(10) \sim O(10^{4})$ times smaller than $F_0$. In other words, the circuit $\hat{U}^{\dagger}$ of only one- and two-qubit gates can fairly evolves $|0 \rangle$ to the targeted MPS with $\chi \gg d$ and $N \gg 1$.

\section{Encoding matrix product state into deep quantum circuit} 

To increase the accuracy systematically, we propose to encode a given MPS $|\psi\rangle$ to a deep quantum circuit that contains multiple layers of MPD's [Fig. \ref{fig-mpd1} (b)]. The number of gates scales linearly with the system size and the number of layers. The encoding algorithm is as following. 
\begin{enumerate}
	\item For the MPS $|\psi_k \rangle$ in the $k$-th iteration (initialized as $|\psi_0 \rangle = |\psi \rangle$), compute the MPS $|\tilde{\psi}_k \rangle$ of $\chi=d$ by optimally truncating the virtual dimensions of $|\psi_k \rangle$;
	\item Compute the MPD $\hat{U}_t$ with $t=\mathcal{D}-k$, which disentangles $|\tilde{\psi}_k \rangle$ to $|0\rangle$, using the method introduced above;
	\item Disentangle $|\psi_k \rangle$ to $|\psi_{k+1} \rangle$ as $|\psi_{k+1} \rangle = \hat{U}_t |\psi_k \rangle$. 
	\item Return the MPD's $\{\hat{U}_t\}$ ($t = 1, 2, \cdots$, $\mathcal{D}$) when the preset number of layers in the circuit is reached, or go back to Step 1 with $|\psi_{k+1} \rangle$. 
\end{enumerate}

\begin{figure}[tbp]
	\centering
	\includegraphics[angle=0,width=1\linewidth]{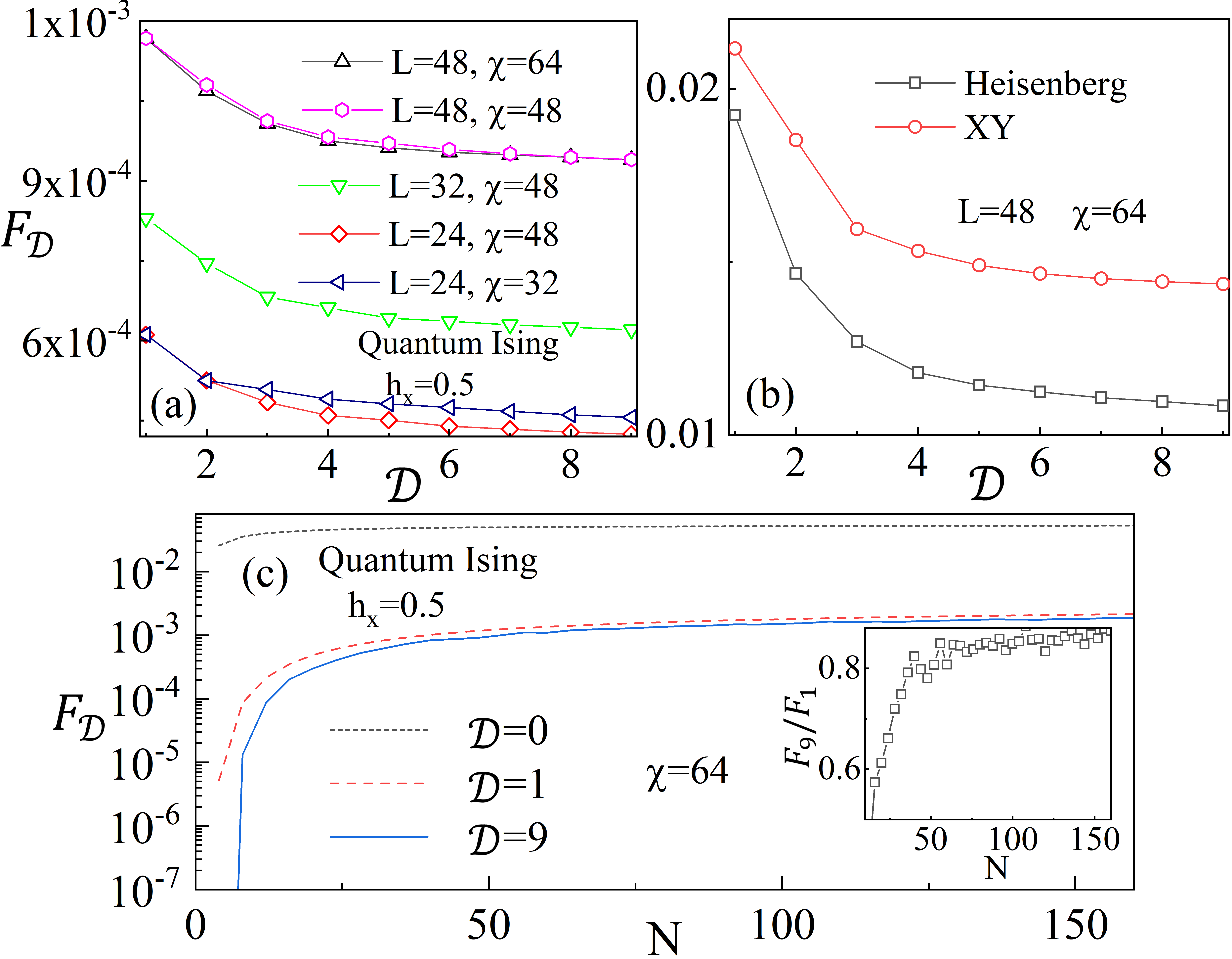}
	\caption{(Color online) The NLF per site $F_{\mathcal{D}}$ [Eq. \ref{eq-fidn}] of the ground states of the 1D (a) quantum Ising model at the critical point, and (b) Heisenberg and XY models, versus the number of layers $\mathcal{D}$ in the quantum circuit. (c) $F_{\mathcal{D}}$ versus the number of qubits $N$ with $\mathcal{D}=0$ (global entanglement), $1$, and $9$ on the transverse Ising model at the critical point. The inset shows $\mathcal{F}_9/\mathcal{F}_1$.}
	\label{fig-layers}
\end{figure}

We testify the encoding algorithm on the ground states of 1D transverse Ising, Heisenberg, and XY models. For the the Heisenberg and XY models, the Hamiltonians are $\hat{H} = \sum_n (\hat{S}^x_n \hat{S}^x_{n+1} + \hat{S}^y_n \hat{S}^y_{n+1} + \hat{S}^z_n \hat{S}^z_{n+1})$ and $\hat{H} = \sum_n (\hat{S}^x_n \hat{S}^x_{n+1} + \hat{S}^y_n \hat{S}^y_{n+1})$, respectively. These Hamiltonians are gapless (rigorously speaking, the true gapless point is in the thermodynamic limit), thus large $\chi$'s of the MPS's are required to carry the entanglement. The NLF between the MPS and the evolved state is defined as 
\begin{equation}\label{eq-fidn}
F_{\mathcal{D}} = -\frac{\ln |\langle \psi | \hat{U}_{\mathcal{D}}^{\dagger} \cdots \hat{U}_2^{\dagger} \hat{U}_1^{\dagger} |0 \rangle|}{N}.
\end{equation}
$F_{\mathcal{D}}$ further decays with $\mathcal{D}$ [Fig. \ref{fig-layers} (a) and (b)]. Note that the largest drop of $F_{\mathcal{D}}$ occurs for $\mathcal{D}=1$, which is reduced by about $O(10^2)$ times compared with $F_0$. For $\mathcal{D} > 1$, relatively large drops of $F_{\mathcal{D}}$ occur for about $\mathcal{D} \leq 4$. With $\mathcal{D}=9$, $F_{\mathcal{D}}$ is reduced by about 20\% - 40\% compared with $F_1$. 

Fig. \ref{fig-layers} (c) shows the NLF's versus the system size $N$ with $\mathcal{D}=0$, $1$, and $9$, on the transverse Ising model at the critical point . The NLF's firstly increase and soon converge as the system becomes larger. The inset shows how the accuracy of the encoding is improved by increasing the number of layers from $\mathcal{D}=1$ to $9$. One can see that $F_9/F_1$ converges to about $0.87$ for large sizes. This indicates that even for the genuine critical systems (in the thermodynamic limit), our scheme robustly prepares the quantum states with high fidelities.

The number for coefficients is significantly compressed by encoding the MPS to the quantum circuit. With $\chi=64$ for instance, the number (per site) for the original MPS $\#(|\psi \rangle) = d\chi^2 = 2^{13}$ is significantly compressed to $\#(\hat{U}) = \mathcal{D} d^4 = 2^7$ for $\mathcal{D} = 8$. Meanwhile, the numbers of both the gates and qubits scale linearly with $N$ and $\mathcal{D}$ [Fig. \ref{fig-mpd1} (b)]. 

Furthermore, by incorporating with the qubit-efficient scheme \cite{HPWS18TNQML}, the number of qubits becomes independent on $N$ and scales only linearly with $\mathcal{D}$. Fig. \ref{fig-mpd1} (c) shows the qubit-efficient scheme of the $\mathcal{D}=2$ circuit with $N=8$. Such an efficient circuit in fact does not gives $N$ entangled qubits, but use only $M=4$ qubits to reproduce the quantum probability distributions (amplitudes) of the $N$-qubit MPS. The trick is to iteratively measure and reuse the last qubit, so that the measurements of this qubit are the same as the measurements on the MPS with $N$ entangled qubits. The first ($M-1$) qubits are to carry the entanglement, thus we have $\chi \leq d^{M-1}$. The total number of qubits in the qubit-efficient scheme equals $\mathcal{D}+2$, which is independent on the number of qubits $N$ in the targeted entangled state. In this way, the circuit is transformed into a more compact form with the same depth but much less qubits, whose number $M$ determines the upper bound of the entanglement entropy as $S \leq (M-1) \ln d$. This makes it possible to realize an MPS on the near-term quantum platforms. For example, for a 20-qubit GHZ state (see some recent progresses in experiments \cite{SXLZ+19GHZ20, OLKS+19CatExp}), which can be written as an MPS with $N=20$ and $d=\chi=2$, the circuit will consist of only three qubits and twenty two-qubit gates. We would like to stress that our deep scheme is not limited to qubits (meaning taking $d=2$). For an MPS of $d$-level qudits with $\chi \gg d$, our scheme can be applied to prepare the MPS's by using only the gates of two ($d$-level) qudits.

\section{Error propagations and complexity in classical simulations} 

The classical computational cost of calculating the MPD's with the encoding algorithm scales only linearly with $\mathcal{D}$. One can see that disentangling the state $|\psi_{k+1} \rangle = \hat{U}_t |\psi_k \rangle$ in Step 3 will essentially increase exponentially the virtual dimensions of $|\psi_k \rangle$ as $\chi d^k$. Therefore, we set an upper bound of the virtual dimensions $\tilde{\chi}$ and truncate if the dimensions exceed $\tilde{\chi}$. Same truncation rule is implemented when calculating $F_{\mathcal{D}}$. The truncation errors are well controlled for $\mathcal{D} \leq \log_d \tilde{\chi}$, same as the time-evolving block decimation algorithm \cite{V04TEBD, V07iTEBD}. 

\begin{figure}[tbp]
	\centering
	\includegraphics[angle=0,width=0.8\linewidth]{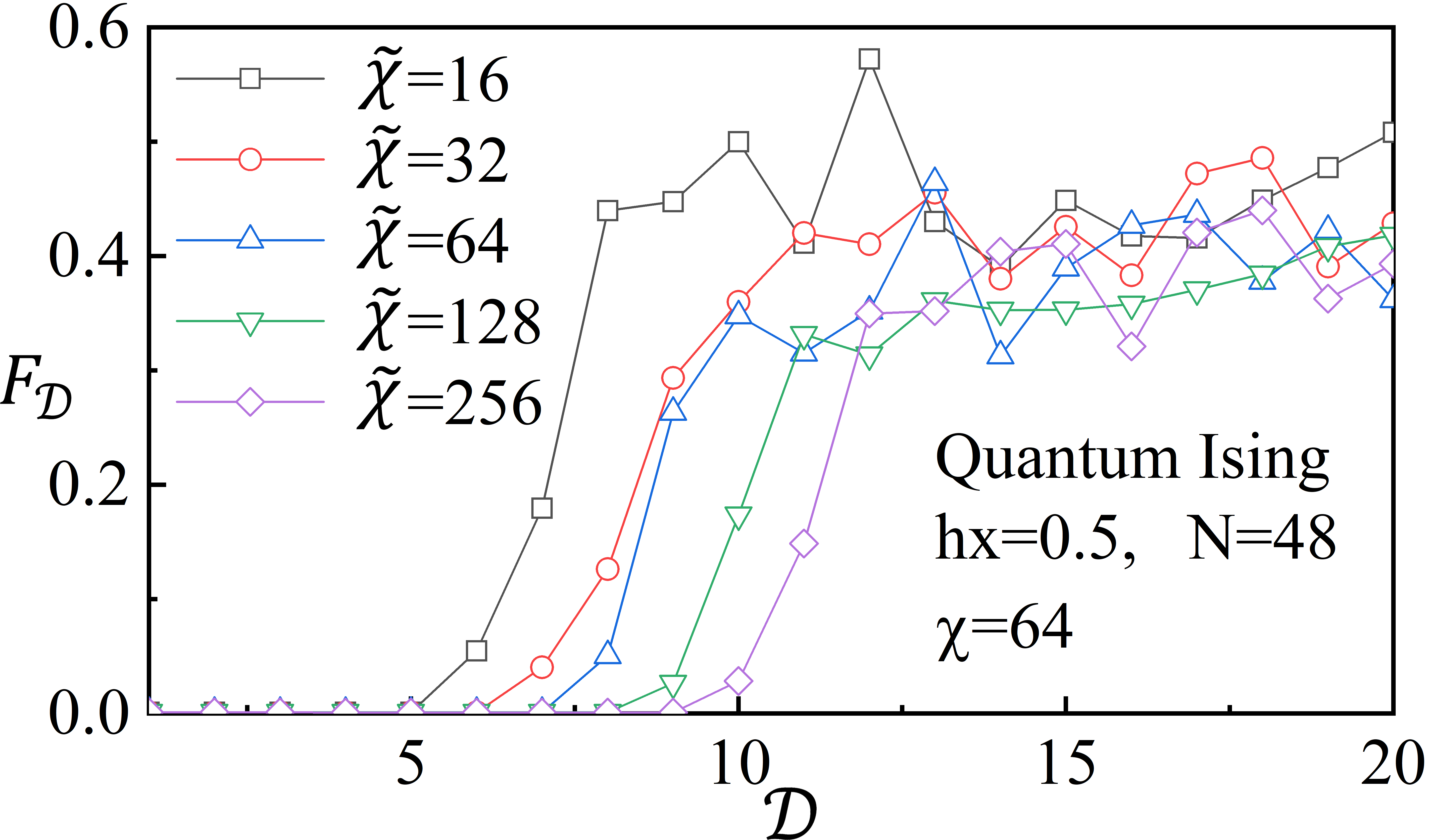}
	\caption{(Color online) The NLF per site $F_{\mathcal{D}}$ of the ground state of 1D quantum Ising model at the critical potin versus the number of layers $\mathcal{D}$ of the quantum circuit. Different dimension cut-offs $\tilde{\chi}$ in the encoding algorithm are taken. The sudden increase of the NLF occurs at $\mathcal{D} = \log_2 \tilde{\chi}$.}
	\label{fig-catastrope}
\end{figure}

For $\mathcal{D} > \log_d \tilde{\chi}$, however, our data show that the NLF suddenly soars (Fig. \ref{fig-catastrope}). This is due to the fact that the truncation errors in the encoding algorithm and those in the evolution by the circuit propagate in opposite directions. In the encoding algorithm, the gates are computed from $\hat{U}_{\mathcal{D}}$ to $\hat{U}_1$. The error accumulates slowly in the same way. For evolution, the gates are acted to $|0\rangle$ from $\hat{U}_1^{\dagger}$ to $\hat{U}_{\mathcal{D}}^{\dagger}$. The virtual dimensions of $|\phi_t \rangle = \hat{U}_t^{\dagger} \cdots \hat{U}_1^{\dagger} |0\rangle$ increases with $t$ as $d^t$ in classical simulations. For $\mathcal{D} \leq \log_d \tilde{\chi}$, there is no truncation error in $|\phi_{\mathcal{D}} \rangle$, thus the error is robustly controlled in the encoding algorithm, and $F_{\mathcal{D}}$ decays with $\mathcal{D}$ as expected. For $\mathcal{D} > \log_d \tilde{\chi}$, truncations have to be implemented. These truncations occur at small $k$'s for the encoding scheme, thus will be propagated to raise the error as soon as the truncations on $|\phi_t$ appear. This makes the classical computational complexity exponentially high, since one needs to keep exponentially large $\tilde{\chi} \sim d^{\mathcal{D}}$ to avoid the error propagations. Note one will not have this issue for quantum computations.

\section{Summary}

Encoding an $N$-qubit MPS with large virtual dimensions to a quantum circuit is an important but extremely challenging task. In this work, we propose an efficient and accurate algorithm that encodes a given MPS into a quantum circuit that consists of only one- and two-qubit gates. We testified our algorithm on several MPS's that describe the ground states of the ``nearly'' gapless Hamiltonians. These MPS's possess large entanglement, thus are difficult to encode using the existing methods. Our data show that the deep quantum circuit constructed by our algorithm can accurately and efficiently evolve a product state to the targeted MPS which might even have large virtual dimensions and/or system size. 

% Note for $d>2$, one should carefully guarantee that $\dim(s_n) \dim(a_{n}) \geq \dim(a_{n-1})$ and $\dim(s_N) \geq \dim(a_{N-1})$ to ensure the orthogonal conditions given in Eqs. (\ref{eq-cond2}) and (\ref{eq-cond3}).

\section*{Acknowledgment} 

SJR is grateful to Jin-Guo Liu for stimulating discussions and useful information. This work was supported by Beijing Natural Science Foundation (Grant No. 1192005 and No. Z180013) and the NSFC (Grant No. 11834014), and by the Academy for Multidisciplinary Studies, Capital Normal University.

\appendix
\setcounter{equation}{0}
\setcounter{figure}{0}
%\setcounter{table}{0}
%\setcounter{page}{1}
%\makeatletter
\renewcommand{\theequation}{A\arabic{equation}}
\renewcommand{\thefigure}{A\arabic{figure}}

\section*{Appendix: Proof of matrix product disentangler being unitary}

\begin{figure}[tb]
	\centering
	\includegraphics[angle=0,width=1\linewidth]{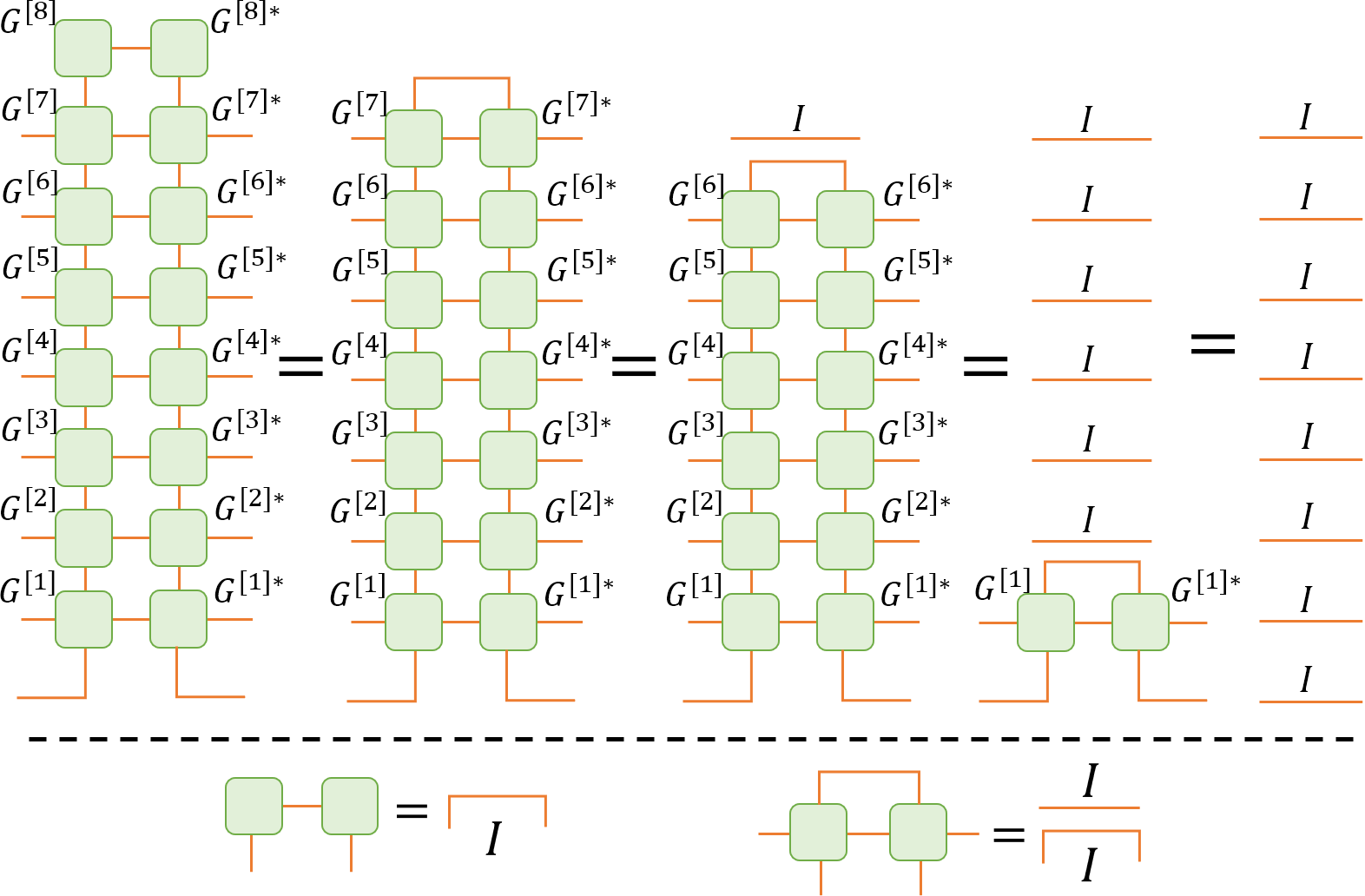}
	\caption{The diagrammatic proof that the matrix product disentangler (MPD) is unitary. $I$ denotes the identity.}
	\label{fig-proof}
\end{figure}

In the Appendix, we give the diagrammatic proof that the matrix product disentangler (MPD) is unitary (Fig. \ref{fig-proof}). The idea is to use the unitary conditions of the orthogonal matrix product state (MPS) and the orthogonal properties in the kernel space. We take the number of qubits $N=8$ as an example. The proof can be readily generalized to any other $N$. 

First, $G^{[8]}$ and $G^{[8]\ast}$ vanish into an identity [illustrated at the bottom of the figure], since
\begin{eqnarray}
&&G^{[8]}=A^{[8]};\nonumber \\
&&\sum_{s_8} A^{[8]}_{s_8, a_{8-1}} A^{[8]\ast}_{s_8, a'_{8-1}} = I_{a_{8-1} a'_{8-1}}. \nonumber
\end{eqnarray}

For the gates corresponding to the tensors in the middle of the MPS ($1<n<N$), they are unitary satisfying $\hat{G}^{[n]\dagger} \hat{G}^{[n]} =I$. The reason is as following. The basis in $\hat{G}^{[n]}$ consist of two parts. One part is the basis given by the tensor of the MPS, which is in fact an isometry due to the orthogonal form of the MPS, i.e.
\begin{eqnarray}
\sum_{s_na_{n}} A^{[n]}_{s_n, a_{n-1} a_{n}} A^{[n]\ast}_{s_n, a'_{n-1} a_{n}}  = I_{a_{n-1} a'_{n-1}}. \nonumber
\end{eqnarray}
Therefore, these basis are orthonormal. The other part contains the basis defined in the kernel space. These basis are orthonormal within themselves by definition, and are orthogonal to the basis in the isometry due to the property of the kernel. Consequently, these gates sequentially result in identities, as shown in the figure below.

For the gate $\hat{G}^{[1]}$, it also contains two parts. one of the basis is the first tensor in the MPS, which is normalized as it is the orthogonal center, i.e.,
\begin{eqnarray}
\sum_{s_1a_1} A^{[1]}_{s_1, a_1} A^{[1]\ast}_{s_1, a_1} = 1. \nonumber
\end{eqnarray}
The rest basis are orthonormal basis in the kernel. Therefore, we have $\hat{G}^{[1]\dagger} \hat{G^{[1]}} =I$. 

Using the orthogonal conditions of the gates, we have $\hat{U}^{\dagger} \hat{U} =I$ as illustrated in the figure below. Since $\hat{U}$ is in fact a ($d^N \times d^N$) square matrix, one readily has $\hat{U} \hat{U}^{\dagger} =I$.

% \bibliographystyle{apsrev4-1}
% \bibliography{ranshiju}

\begin{thebibliography}{55}%
\makeatletter
\providecommand \@ifxundefined [1]{%
 \@ifx{#1\undefined}
}%
\providecommand \@ifnum [1]{%
 \ifnum #1\expandafter \@firstoftwo
 \else \expandafter \@secondoftwo
 \fi
}%
\providecommand \@ifx [1]{%
 \ifx #1\expandafter \@firstoftwo
 \else \expandafter \@secondoftwo
 \fi
}%
\providecommand \natexlab [1]{#1}%
\providecommand \enquote  [1]{``#1''}%
\providecommand \bibnamefont  [1]{#1}%
\providecommand \bibfnamefont [1]{#1}%
\providecommand \citenamefont [1]{#1}%
\providecommand \href@noop [0]{\@secondoftwo}%
\providecommand \href [0]{\begingroup \@sanitize@url \@href}%
\providecommand \@href[1]{\@@startlink{#1}\@@href}%
\providecommand \@@href[1]{\endgroup#1\@@endlink}%
\providecommand \@sanitize@url [0]{\catcode `\\12\catcode `\$12\catcode
  `\&12\catcode `\#12\catcode `\^12\catcode `\_12\catcode `\%12\relax}%
\providecommand \@@startlink[1]{}%
\providecommand \@@endlink[0]{}%
\providecommand \url  [0]{\begingroup\@sanitize@url \@url }%
\providecommand \@url [1]{\endgroup\@href {#1}{\urlprefix }}%
\providecommand \urlprefix  [0]{URL }%
\providecommand \Eprint [0]{\href }%
\providecommand \doibase [0]{http://dx.doi.org/}%
\providecommand \selectlanguage [0]{\@gobble}%
\providecommand \bibinfo  [0]{\@secondoftwo}%
\providecommand \bibfield  [0]{\@secondoftwo}%
\providecommand \translation [1]{[#1]}%
\providecommand \BibitemOpen [0]{}%
\providecommand \bibitemStop [0]{}%
\providecommand \bibitemNoStop [0]{.\EOS\space}%
\providecommand \EOS [0]{\spacefactor3000\relax}%
\providecommand \BibitemShut  [1]{\csname bibitem#1\endcsname}%
\let\auto@bib@innerbib\@empty
%</preamble>
\bibitem [{\citenamefont {White}(1992)}]{W92DMRG}%
  \BibitemOpen
  \bibfield  {author} {\bibinfo {author} {\bibfnamefont {Steven~R.}\
  \bibnamefont {White}},\ }\bibfield  {title} {\enquote {\bibinfo {title}
  {Density matrix formulation for quantum renormalization groups},}\ }\href
  {\doibase 10.1103/PhysRevLett.69.2863} {\bibfield  {journal} {\bibinfo
  {journal} {Phys. Rev. Lett.}\ }\textbf {\bibinfo {volume} {69}},\ \bibinfo
  {pages} {2863--2866} (\bibinfo {year} {1992})}\BibitemShut {NoStop}%
\bibitem [{\citenamefont {White}(1993)}]{W93DMRG}%
  \BibitemOpen
  \bibfield  {author} {\bibinfo {author} {\bibfnamefont {Steven~R.}\
  \bibnamefont {White}},\ }\bibfield  {title} {\enquote {\bibinfo {title}
  {{Density-matrix algorithms for quantum renormalization groups}},}\ }\href
  {\doibase 10.1103/PhysRevB.48.10345} {\bibfield  {journal} {\bibinfo
  {journal} {Phys. Rev. B}\ }\textbf {\bibinfo {volume} {48}},\ \bibinfo
  {pages} {10345--10356} (\bibinfo {year} {1993})}\BibitemShut {NoStop}%
\bibitem [{\citenamefont {Dukelsky}\ \emph {et~al.}(1998)\citenamefont
  {Dukelsky}, \citenamefont {Mart\'in-Del\'{g}ado}, \citenamefont {Nishino},\
  and\ \citenamefont {Sierra}}]{DMNS98MPS}%
  \BibitemOpen
  \bibfield  {author} {\bibinfo {author} {\bibfnamefont {Jorge}\ \bibnamefont
  {Dukelsky}}, \bibinfo {author} {\bibfnamefont {Miguel~A.}\ \bibnamefont
  {Mart\'in-Del\'{g}ado}}, \bibinfo {author} {\bibfnamefont {Tomotoshi}\
  \bibnamefont {Nishino}}, \ and\ \bibinfo {author} {\bibfnamefont
  {Germ{\'{a}}n}\ \bibnamefont {Sierra}},\ }\bibfield  {title} {\enquote
  {\bibinfo {title} {{Equivalence of the variational matrix product method and
  the density matrix renormalization group applied to spin chains}},}\ }\href
  {\doibase 10.1209/epl/i1998-00381-x} {\bibfield  {journal} {\bibinfo
  {journal} {Europhys. Lett.}\ }\textbf {\bibinfo {volume} {43}},\ \bibinfo
  {pages} {457} (\bibinfo {year} {1998})}\BibitemShut {NoStop}%
\bibitem [{\citenamefont {Moukouri}\ and\ \citenamefont
  {Caron}(1996)}]{MC96FTDMRG}%
  \BibitemOpen
  \bibfield  {author} {\bibinfo {author} {\bibfnamefont {S.}~\bibnamefont
  {Moukouri}}\ and\ \bibinfo {author} {\bibfnamefont {L.~G.}\ \bibnamefont
  {Caron}},\ }\bibfield  {title} {\enquote {\bibinfo {title} {Thermodynamic
  density matrix renormalization group study of the magnetic susceptibility of
  half-integer quantum spin chains},}\ }\href {\doibase
  10.1103/PhysRevLett.77.4640} {\bibfield  {journal} {\bibinfo  {journal}
  {Phys. Rev. Lett.}\ }\textbf {\bibinfo {volume} {77}},\ \bibinfo {pages}
  {4640--4643} (\bibinfo {year} {1996})}\BibitemShut {NoStop}%
\bibitem [{\citenamefont {Wang}\ and\ \citenamefont {Xiang}(1997)}]{WX97TMRG}%
  \BibitemOpen
  \bibfield  {author} {\bibinfo {author} {\bibfnamefont {Xiaoqun}\ \bibnamefont
  {Wang}}\ and\ \bibinfo {author} {\bibfnamefont {Tao}\ \bibnamefont {Xiang}},\
  }\bibfield  {title} {\enquote {\bibinfo {title} {Transfer-matrix
  density-matrix renormalization-group theory for thermodynamics of
  one-dimensional quantum systems},}\ }\href {\doibase
  10.1103/PhysRevB.56.5061} {\bibfield  {journal} {\bibinfo  {journal} {Phys.
  Rev. B}\ }\textbf {\bibinfo {volume} {56}},\ \bibinfo {pages} {5061--5064}
  (\bibinfo {year} {1997})}\BibitemShut {NoStop}%
\bibitem [{\citenamefont {Bartel}\ \emph {et~al.}(2003)\citenamefont {Bartel},
  \citenamefont {Schadschneider},\ and\ \citenamefont
  {Zittartz}}]{BSZ03exciteMPS}%
  \BibitemOpen
  \bibfield  {author} {\bibinfo {author} {\bibfnamefont {E.}~\bibnamefont
  {Bartel}}, \bibinfo {author} {\bibfnamefont {A.}~\bibnamefont
  {Schadschneider}}, \ and\ \bibinfo {author} {\bibfnamefont {J.}~\bibnamefont
  {Zittartz}},\ }\bibfield  {title} {\enquote {\bibinfo {title} {Excitations of
  anisotropic spin-1 chains with matrix product ground state},}\ }\href
  {\doibase 10.1140/epjb/e2003-00025-7} {\bibfield  {journal} {\bibinfo
  {journal} {J. Eur. Phys. J. B}\ }\textbf {\bibinfo {volume} {31}},\ \bibinfo
  {pages} {209--216} (\bibinfo {year} {2003})}\BibitemShut {NoStop}%
\bibitem [{\citenamefont {Chung}\ and\ \citenamefont
  {Wang}(2009)}]{CW09EntPurt}%
  \BibitemOpen
  \bibfield  {author} {\bibinfo {author} {\bibfnamefont {Sung~Gong}\
  \bibnamefont {Chung}}\ and\ \bibinfo {author} {\bibfnamefont {Lihua}\
  \bibnamefont {Wang}},\ }\bibfield  {title} {\enquote {\bibinfo {title}
  {Entanglement perturbation theory for the elementary excitation in one
  dimension},}\ }\href {\doibase 10.1016/j.physleta.2009.04.038} {\bibfield
  {journal} {\bibinfo  {journal} {Phys. Lett. A}\ }\textbf {\bibinfo {volume}
  {373}},\ \bibinfo {pages} {2277 -- 2280} (\bibinfo {year}
  {2009})}\BibitemShut {NoStop}%
\bibitem [{\citenamefont {Haegeman}\ \emph {et~al.}(2012)\citenamefont
  {Haegeman}, \citenamefont {Pirvu}, \citenamefont {Weir}, \citenamefont
  {Cirac}, \citenamefont {Osborne}, \citenamefont {Verschelde},\ and\
  \citenamefont {Verstraete}}]{HPWC+12MPSexcitations}%
  \BibitemOpen
  \bibfield  {author} {\bibinfo {author} {\bibfnamefont {Jutho}\ \bibnamefont
  {Haegeman}}, \bibinfo {author} {\bibfnamefont {Bogdan}\ \bibnamefont
  {Pirvu}}, \bibinfo {author} {\bibfnamefont {David~J.}\ \bibnamefont {Weir}},
  \bibinfo {author} {\bibfnamefont {J.~Ignacio}\ \bibnamefont {Cirac}},
  \bibinfo {author} {\bibfnamefont {Tobias~J.}\ \bibnamefont {Osborne}},
  \bibinfo {author} {\bibfnamefont {Henri}\ \bibnamefont {Verschelde}}, \ and\
  \bibinfo {author} {\bibfnamefont {Frank}\ \bibnamefont {Verstraete}},\
  }\bibfield  {title} {\enquote {\bibinfo {title} {Variational matrix product
  ansatz for dispersion relations},}\ }\href {\doibase
  10.1103/PhysRevB.85.100408} {\bibfield  {journal} {\bibinfo  {journal} {Phys.
  Rev. B}\ }\textbf {\bibinfo {volume} {85}},\ \bibinfo {pages} {100408(R)}
  (\bibinfo {year} {2012})}\BibitemShut {NoStop}%
\bibitem [{\citenamefont {Vidal}(2004)}]{V04TEBD}%
  \BibitemOpen
  \bibfield  {author} {\bibinfo {author} {\bibfnamefont {Guifr\'e}\
  \bibnamefont {Vidal}},\ }\bibfield  {title} {\enquote {\bibinfo {title}
  {Efficient simulation of one-dimensional quantum many-body systems},}\ }\href
  {\doibase 10.1103/PhysRevLett.93.040502} {\bibfield  {journal} {\bibinfo
  {journal} {Phys. Rev. Lett.}\ }\textbf {\bibinfo {volume} {93}},\ \bibinfo
  {pages} {040502} (\bibinfo {year} {2004})}\BibitemShut {NoStop}%
\bibitem [{\citenamefont {Verstraete}\ \emph
  {et~al.}(2004{\natexlab{a}})\citenamefont {Verstraete}, \citenamefont
  {Garc\'ia-Ripoll},\ and\ \citenamefont {Cirac}}]{VGC04MPDO}%
  \BibitemOpen
  \bibfield  {author} {\bibinfo {author} {\bibfnamefont {Frank}\ \bibnamefont
  {Verstraete}}, \bibinfo {author} {\bibfnamefont {Juan~J.}\ \bibnamefont
  {Garc\'ia-Ripoll}}, \ and\ \bibinfo {author} {\bibfnamefont {J.~Ignacio}\
  \bibnamefont {Cirac}},\ }\bibfield  {title} {\enquote {\bibinfo {title}
  {{Matrix Product Density Operators: Simulation of Finite-Temperature and
  Dissipative Systems}},}\ }\href {\doibase 10.1103/PhysRevLett.93.207204}
  {\bibfield  {journal} {\bibinfo  {journal} {Phys. Rev. Lett.}\ }\textbf
  {\bibinfo {volume} {93}},\ \bibinfo {pages} {207204} (\bibinfo {year}
  {2004}{\natexlab{a}})}\BibitemShut {NoStop}%
\bibitem [{\citenamefont {Verstraete}\ and\ \citenamefont
  {Cirac}(2006)}]{VC06MPSFaithfully}%
  \BibitemOpen
  \bibfield  {author} {\bibinfo {author} {\bibfnamefont {F.}~\bibnamefont
  {Verstraete}}\ and\ \bibinfo {author} {\bibfnamefont {J.~I.}\ \bibnamefont
  {Cirac}},\ }\bibfield  {title} {\enquote {\bibinfo {title} {Matrix product
  states represent ground states faithfully},}\ }\href {\doibase
  10.1103/PhysRevB.73.094423} {\bibfield  {journal} {\bibinfo  {journal} {Phys.
  Rev. B}\ }\textbf {\bibinfo {volume} {73}},\ \bibinfo {pages} {094423}
  (\bibinfo {year} {2006})}\BibitemShut {NoStop}%
\bibitem [{\citenamefont {Vidal}(2007)}]{V07iTEBD}%
  \BibitemOpen
  \bibfield  {author} {\bibinfo {author} {\bibfnamefont {G.}~\bibnamefont
  {Vidal}},\ }\bibfield  {title} {\enquote {\bibinfo {title} {Classical
  simulation of infinite-size quantum lattice systems in one spatial
  dimension},}\ }\href {\doibase 10.1103/PhysRevLett.98.070201} {\bibfield
  {journal} {\bibinfo  {journal} {Phys. Rev. Lett.}\ }\textbf {\bibinfo
  {volume} {98}},\ \bibinfo {pages} {070201} (\bibinfo {year}
  {2007})}\BibitemShut {NoStop}%
\bibitem [{\citenamefont {Li}\ \emph {et~al.}(2011)\citenamefont {Li},
  \citenamefont {Ran}, \citenamefont {Gong}, \citenamefont {Zhao},
  \citenamefont {Xi}, \citenamefont {Ye},\ and\ \citenamefont
  {Su}}]{LRGZXY+11LTRG}%
  \BibitemOpen
  \bibfield  {author} {\bibinfo {author} {\bibfnamefont {Wei}\ \bibnamefont
  {Li}}, \bibinfo {author} {\bibfnamefont {Shi-Ju}\ \bibnamefont {Ran}},
  \bibinfo {author} {\bibfnamefont {Shou-Shu}\ \bibnamefont {Gong}}, \bibinfo
  {author} {\bibfnamefont {Yang}\ \bibnamefont {Zhao}}, \bibinfo {author}
  {\bibfnamefont {Bin}\ \bibnamefont {Xi}}, \bibinfo {author} {\bibfnamefont
  {Fei}\ \bibnamefont {Ye}}, \ and\ \bibinfo {author} {\bibfnamefont {Gang}\
  \bibnamefont {Su}},\ }\bibfield  {title} {\enquote {\bibinfo {title}
  {Linearized tensor renormalization group algorithm for the calculation of
  thermodynamic properties of quantum lattice models},}\ }\href {\doibase
  10.1103/PhysRevLett.106.127202} {\bibfield  {journal} {\bibinfo  {journal}
  {Phys. Rev. Lett.}\ }\textbf {\bibinfo {volume} {106}},\ \bibinfo {pages}
  {127202} (\bibinfo {year} {2011})}\BibitemShut {NoStop}%
\bibitem [{\citenamefont {Johnson}\ \emph {et~al.}(2010)\citenamefont
  {Johnson}, \citenamefont {Clark},\ and\ \citenamefont
  {Jaksch}}]{JCJ10MPSstat}%
  \BibitemOpen
  \bibfield  {author} {\bibinfo {author} {\bibfnamefont {T.~H.}\ \bibnamefont
  {Johnson}}, \bibinfo {author} {\bibfnamefont {S.~R.}\ \bibnamefont {Clark}},
  \ and\ \bibinfo {author} {\bibfnamefont {D.}~\bibnamefont {Jaksch}},\
  }\bibfield  {title} {\enquote {\bibinfo {title} {Dynamical simulations of
  classical stochastic systems using matrix product states},}\ }\href {\doibase
  10.1103/PhysRevE.82.036702} {\bibfield  {journal} {\bibinfo  {journal} {Phys.
  Rev. E}\ }\textbf {\bibinfo {volume} {82}},\ \bibinfo {pages} {036702}
  (\bibinfo {year} {2010})}\BibitemShut {NoStop}%
\bibitem [{\citenamefont {Prosen}\ and\ \citenamefont
  {{\v{Z}}nidari{\v{c}}}(2009)}]{PZ09MPSsteady}%
  \BibitemOpen
  \bibfield  {author} {\bibinfo {author} {\bibfnamefont {Toma{\v{z}}}\
  \bibnamefont {Prosen}}\ and\ \bibinfo {author} {\bibfnamefont {Marko}\
  \bibnamefont {{\v{Z}}nidari{\v{c}}}},\ }\bibfield  {title} {\enquote
  {\bibinfo {title} {Matrix product simulations of non-equilibrium steady
  states of quantum spin chains},}\ }\href {\doibase
  10.1088/1742-5468/2009/02/p02035} {\bibfield  {journal} {\bibinfo  {journal}
  {J. Stat. Mech.}\ }\textbf {\bibinfo {volume} {2009}},\ \bibinfo {pages}
  {P02035} (\bibinfo {year} {2009})}\BibitemShut {NoStop}%
\bibitem [{\citenamefont {Wolf}\ \emph {et~al.}(2014)\citenamefont {Wolf},
  \citenamefont {McCulloch},\ and\ \citenamefont
  {Schollw\"ock}}]{WMIS14MPSDMFT}%
  \BibitemOpen
  \bibfield  {author} {\bibinfo {author} {\bibfnamefont {F.~Alexander}\
  \bibnamefont {Wolf}}, \bibinfo {author} {\bibfnamefont {Ian~P.}\ \bibnamefont
  {McCulloch}}, \ and\ \bibinfo {author} {\bibfnamefont {Ulrich}\ \bibnamefont
  {Schollw\"ock}},\ }\bibfield  {title} {\enquote {\bibinfo {title} {Solving
  nonequilibrium dynamical mean-field theory using matrix product states},}\
  }\href {\doibase 10.1103/PhysRevB.90.235131} {\bibfield  {journal} {\bibinfo
  {journal} {Phys. Rev. B}\ }\textbf {\bibinfo {volume} {90}},\ \bibinfo
  {pages} {235131} (\bibinfo {year} {2014})}\BibitemShut {NoStop}%
\bibitem [{\citenamefont {Schr\"oder}\ and\ \citenamefont
  {Chin}(2016)}]{SC16MPSopen}%
  \BibitemOpen
  \bibfield  {author} {\bibinfo {author} {\bibfnamefont {Florian A. Y.~N.}\
  \bibnamefont {Schr\"oder}}\ and\ \bibinfo {author} {\bibfnamefont {Alex~W.}\
  \bibnamefont {Chin}},\ }\bibfield  {title} {\enquote {\bibinfo {title}
  {Simulating open quantum dynamics with time-dependent variational matrix
  product states: Towards microscopic correlation of environment dynamics and
  reduced system evolution},}\ }\href {\doibase 10.1103/PhysRevB.93.075105}
  {\bibfield  {journal} {\bibinfo  {journal} {Phys. Rev. B}\ }\textbf {\bibinfo
  {volume} {93}},\ \bibinfo {pages} {075105} (\bibinfo {year}
  {2016})}\BibitemShut {NoStop}%
\bibitem [{\citenamefont {Jaschke}\ \emph {et~al.}(2018)\citenamefont
  {Jaschke}, \citenamefont {Montangero},\ and\ \citenamefont
  {Carr}}]{JMC18MPSopen}%
  \BibitemOpen
  \bibfield  {author} {\bibinfo {author} {\bibfnamefont {Daniel}\ \bibnamefont
  {Jaschke}}, \bibinfo {author} {\bibfnamefont {Simone}\ \bibnamefont
  {Montangero}}, \ and\ \bibinfo {author} {\bibfnamefont {Lincoln~D.}\
  \bibnamefont {Carr}},\ }\bibfield  {title} {\enquote {\bibinfo {title}
  {One-dimensional many-body entangled open quantum systems with tensor network
  methods},}\ }\href {\doibase 10.1088/2058-9565/aae724} {\bibfield  {journal}
  {\bibinfo  {journal} {Quantum Sci. Technol.}\ }\textbf {\bibinfo {volume}
  {4}},\ \bibinfo {pages} {013001} (\bibinfo {year} {2018})}\BibitemShut
  {NoStop}%
\bibitem [{\citenamefont {Cirac}\ and\ \citenamefont
  {Sierra}(2010)}]{CS10critical}%
  \BibitemOpen
  \bibfield  {author} {\bibinfo {author} {\bibfnamefont {J.~Ignacio}\
  \bibnamefont {Cirac}}\ and\ \bibinfo {author} {\bibfnamefont {Germ\'{a}n}\
  \bibnamefont {Sierra}},\ }\bibfield  {title} {\enquote {\bibinfo {title}
  {Infinite matrix product states, conformal field theory, and the
  haldane-shastry model},}\ }\href {\doibase 10.1103/PhysRevB.81.104431}
  {\bibfield  {journal} {\bibinfo  {journal} {Phys. Rev. B}\ }\textbf {\bibinfo
  {volume} {81}},\ \bibinfo {pages} {104431} (\bibinfo {year}
  {2010})}\BibitemShut {NoStop}%
\bibitem [{\citenamefont {Verstraete}\ and\ \citenamefont
  {Cirac}(2010)}]{VC10cMPS}%
  \BibitemOpen
  \bibfield  {author} {\bibinfo {author} {\bibfnamefont {F.}~\bibnamefont
  {Verstraete}}\ and\ \bibinfo {author} {\bibfnamefont {J.~I.}\ \bibnamefont
  {Cirac}},\ }\bibfield  {title} {\enquote {\bibinfo {title} {Continuous matrix
  product states for quantum fields},}\ }\href {\doibase
  10.1103/PhysRevLett.104.190405} {\bibfield  {journal} {\bibinfo  {journal}
  {Phys. Rev. Lett.}\ }\textbf {\bibinfo {volume} {104}},\ \bibinfo {pages}
  {190405} (\bibinfo {year} {2010})}\BibitemShut {NoStop}%
\bibitem [{\citenamefont {Haegeman}\ \emph {et~al.}(2013)\citenamefont
  {Haegeman}, \citenamefont {Cirac}, \citenamefont {Osborne},\ and\
  \citenamefont {Verstraete}}]{HCOV13cMPS}%
  \BibitemOpen
  \bibfield  {author} {\bibinfo {author} {\bibfnamefont {Jutho}\ \bibnamefont
  {Haegeman}}, \bibinfo {author} {\bibfnamefont {J.~Ignacio}\ \bibnamefont
  {Cirac}}, \bibinfo {author} {\bibfnamefont {Tobias~J.}\ \bibnamefont
  {Osborne}}, \ and\ \bibinfo {author} {\bibfnamefont {Frank}\ \bibnamefont
  {Verstraete}},\ }\bibfield  {title} {\enquote {\bibinfo {title} {Calculus of
  continuous matrix product states},}\ }\href {\doibase
  10.1103/PhysRevB.88.085118} {\bibfield  {journal} {\bibinfo  {journal} {Phys.
  Rev. B}\ }\textbf {\bibinfo {volume} {88}},\ \bibinfo {pages} {085118}
  (\bibinfo {year} {2013})}\BibitemShut {NoStop}%
\bibitem [{\citenamefont {Milsted}\ \emph {et~al.}(2013)\citenamefont
  {Milsted}, \citenamefont {Haegeman},\ and\ \citenamefont
  {Osborne}}]{MHO13MPSfield}%
  \BibitemOpen
  \bibfield  {author} {\bibinfo {author} {\bibfnamefont {Ashley}\ \bibnamefont
  {Milsted}}, \bibinfo {author} {\bibfnamefont {Jutho}\ \bibnamefont
  {Haegeman}}, \ and\ \bibinfo {author} {\bibfnamefont {Tobias~J.}\
  \bibnamefont {Osborne}},\ }\bibfield  {title} {\enquote {\bibinfo {title}
  {Matrix product states and variational methods applied to critical quantum
  field theory},}\ }\href {\doibase 10.1103/PhysRevD.88.085030} {\bibfield
  {journal} {\bibinfo  {journal} {Phys. Rev. D}\ }\textbf {\bibinfo {volume}
  {88}},\ \bibinfo {pages} {085030} (\bibinfo {year} {2013})}\BibitemShut
  {NoStop}%
\bibitem [{\citenamefont {Steffens}\ \emph {et~al.}(2014)\citenamefont
  {Steffens}, \citenamefont {Riofr\'io}, \citenamefont {H\"ubener},\ and\
  \citenamefont {Eisert}}]{SRHE14fieldcMPS}%
  \BibitemOpen
  \bibfield  {author} {\bibinfo {author} {\bibfnamefont {A.}~\bibnamefont
  {Steffens}}, \bibinfo {author} {\bibfnamefont {C.~A.}\ \bibnamefont
  {Riofr\'io}}, \bibinfo {author} {\bibfnamefont {R.}~\bibnamefont
  {H\"ubener}}, \ and\ \bibinfo {author} {\bibfnamefont {J.}~\bibnamefont
  {Eisert}},\ }\bibfield  {title} {\enquote {\bibinfo {title} {Quantum field
  tomography},}\ }\href {\doibase 10.1088/1367-2630/16/12/123010} {\bibfield
  {journal} {\bibinfo  {journal} {New J. Phys.}\ }\textbf {\bibinfo {volume}
  {16}},\ \bibinfo {pages} {123010} (\bibinfo {year} {2014})}\BibitemShut
  {NoStop}%
\bibitem [{\citenamefont {Rinc\'on}\ \emph {et~al.}(2015)\citenamefont
  {Rinc\'on}, \citenamefont {Ganahl},\ and\ \citenamefont {Vidal}}]{RGV15cMPS}%
  \BibitemOpen
  \bibfield  {author} {\bibinfo {author} {\bibfnamefont {Juli\'an}\
  \bibnamefont {Rinc\'on}}, \bibinfo {author} {\bibfnamefont {Martin}\
  \bibnamefont {Ganahl}}, \ and\ \bibinfo {author} {\bibfnamefont {Guifre}\
  \bibnamefont {Vidal}},\ }\bibfield  {title} {\enquote {\bibinfo {title}
  {Lieb-liniger model with exponentially decaying interactions: A continuous
  matrix product state study},}\ }\href {\doibase 10.1103/PhysRevB.92.115107}
  {\bibfield  {journal} {\bibinfo  {journal} {Phys. Rev. B}\ }\textbf {\bibinfo
  {volume} {92}},\ \bibinfo {pages} {115107} (\bibinfo {year}
  {2015})}\BibitemShut {NoStop}%
\bibitem [{\citenamefont {Stoudenmire}\ and\ \citenamefont
  {Schwab}(2016)}]{SS16TNML}%
  \BibitemOpen
  \bibfield  {author} {\bibinfo {author} {\bibfnamefont {Edwin}\ \bibnamefont
  {Stoudenmire}}\ and\ \bibinfo {author} {\bibfnamefont {David~J}\ \bibnamefont
  {Schwab}},\ }\bibfield  {title} {\enquote {\bibinfo {title} {Supervised
  learning with tensor networks},}\ }in\ \href
  {http://papers.nips.cc/paper/6211-supervised-learning-with-tensor-networks.pdf}
  {\emph {\bibinfo {booktitle} {Advances in Neural Information Processing
  Systems 29}}},\ \bibinfo {editor} {edited by\ \bibinfo {editor}
  {\bibfnamefont {D.~D.}\ \bibnamefont {Lee}}, \bibinfo {editor} {\bibfnamefont
  {M.}~\bibnamefont {Sugiyama}}, \bibinfo {editor} {\bibfnamefont {U.~V.}\
  \bibnamefont {Luxburg}}, \bibinfo {editor} {\bibfnamefont {I.}~\bibnamefont
  {Guyon}}, \ and\ \bibinfo {editor} {\bibfnamefont {R.}~\bibnamefont
  {Garnett}}}\ (\bibinfo  {publisher} {Curran Associates, Inc.},\ \bibinfo
  {year} {2016})\ pp.\ \bibinfo {pages} {4799--4807}\BibitemShut {NoStop}%
\bibitem [{\citenamefont {Han}\ \emph {et~al.}(2018)\citenamefont {Han},
  \citenamefont {Wang}, \citenamefont {Fan}, \citenamefont {Wang},\ and\
  \citenamefont {Zhang}}]{HWFWZ17MPSML}%
  \BibitemOpen
  \bibfield  {author} {\bibinfo {author} {\bibfnamefont {Zhao-Yu}\ \bibnamefont
  {Han}}, \bibinfo {author} {\bibfnamefont {Jun}\ \bibnamefont {Wang}},
  \bibinfo {author} {\bibfnamefont {Heng}\ \bibnamefont {Fan}}, \bibinfo
  {author} {\bibfnamefont {Lei}\ \bibnamefont {Wang}}, \ and\ \bibinfo {author}
  {\bibfnamefont {Pan}\ \bibnamefont {Zhang}},\ }\bibfield  {title} {\enquote
  {\bibinfo {title} {Unsupervised generative modeling using matrix product
  states},}\ }\href {\doibase 10.1103/PhysRevX.8.031012} {\bibfield  {journal}
  {\bibinfo  {journal} {Phys. Rev. X}\ }\textbf {\bibinfo {volume} {8}},\
  \bibinfo {pages} {031012} (\bibinfo {year} {2018})}\BibitemShut {NoStop}%
\bibitem [{\citenamefont {Chen}\ \emph {et~al.}(2018)\citenamefont {Chen},
  \citenamefont {Cheng}, \citenamefont {Xie}, \citenamefont {Wang},\ and\
  \citenamefont {Xiang}}]{CCXWX18MPSbolzmann}%
  \BibitemOpen
  \bibfield  {author} {\bibinfo {author} {\bibfnamefont {Jing}\ \bibnamefont
  {Chen}}, \bibinfo {author} {\bibfnamefont {Song}\ \bibnamefont {Cheng}},
  \bibinfo {author} {\bibfnamefont {Haidong}\ \bibnamefont {Xie}}, \bibinfo
  {author} {\bibfnamefont {Lei}\ \bibnamefont {Wang}}, \ and\ \bibinfo {author}
  {\bibfnamefont {Tao}\ \bibnamefont {Xiang}},\ }\bibfield  {title} {\enquote
  {\bibinfo {title} {Equivalence of restricted boltzmann machines and tensor
  network states},}\ }\href {\doibase 10.1103/PhysRevB.97.085104} {\bibfield
  {journal} {\bibinfo  {journal} {Phys. Rev. B}\ }\textbf {\bibinfo {volume}
  {97}},\ \bibinfo {pages} {085104} (\bibinfo {year} {2018})}\BibitemShut
  {NoStop}%
\bibitem [{\citenamefont {Huggins}\ \emph {et~al.}(2019)\citenamefont
  {Huggins}, \citenamefont {Patil}, \citenamefont {Mitchell}, \citenamefont
  {Whaley},\ and\ \citenamefont {Stoudenmire}}]{HPWS18TNQML}%
  \BibitemOpen
  \bibfield  {author} {\bibinfo {author} {\bibfnamefont {William}\ \bibnamefont
  {Huggins}}, \bibinfo {author} {\bibfnamefont {Piyush}\ \bibnamefont {Patil}},
  \bibinfo {author} {\bibfnamefont {Bradley}\ \bibnamefont {Mitchell}},
  \bibinfo {author} {\bibfnamefont {K~Birgitta}\ \bibnamefont {Whaley}}, \ and\
  \bibinfo {author} {\bibfnamefont {E~Miles}\ \bibnamefont {Stoudenmire}},\
  }\bibfield  {title} {\enquote {\bibinfo {title} {Towards quantum machine
  learning with tensor networks},}\ }\href {\doibase 10.1088/2058-9565/aaea94}
  {\bibfield  {journal} {\bibinfo  {journal} {Quantum Sci. Technol.}\ }\textbf
  {\bibinfo {volume} {4}},\ \bibinfo {pages} {024001} (\bibinfo {year}
  {2019})}\BibitemShut {NoStop}%
\bibitem [{\citenamefont {Sun}\ \emph {et~al.}(2019)\citenamefont {Sun},
  \citenamefont {Peng}, \citenamefont {Liu}, \citenamefont {Ran},\ and\
  \citenamefont {Su}}]{SPLRS19GTNC}%
  \BibitemOpen
  \bibfield  {author} {\bibinfo {author} {\bibfnamefont {Zheng-Zhi}\
  \bibnamefont {Sun}}, \bibinfo {author} {\bibfnamefont {Cheng}\ \bibnamefont
  {Peng}}, \bibinfo {author} {\bibfnamefont {Ding}\ \bibnamefont {Liu}},
  \bibinfo {author} {\bibfnamefont {Shi-Ju}\ \bibnamefont {Ran}}, \ and\
  \bibinfo {author} {\bibfnamefont {Gang}\ \bibnamefont {Su}},\ }\bibfield
  {title} {\enquote {\bibinfo {title} {Generative tensor network classification
  model for supervised machine learning},}\ }\href@noop {} {\  (\bibinfo {year}
  {2019})},\ \Eprint {http://arxiv.org/abs/arXiv:1903.10742} {arXiv:1903.10742}
  \BibitemShut {NoStop}%
\bibitem [{\citenamefont {Verstraete}\ \emph
  {et~al.}(2004{\natexlab{b}})\citenamefont {Verstraete}, \citenamefont
  {Porras},\ and\ \citenamefont {Cirac}}]{VPC04DMRGQinfo}%
  \BibitemOpen
  \bibfield  {author} {\bibinfo {author} {\bibfnamefont {Frank}\ \bibnamefont
  {Verstraete}}, \bibinfo {author} {\bibfnamefont {Diego}\ \bibnamefont
  {Porras}}, \ and\ \bibinfo {author} {\bibfnamefont {J.~Ignacio}\ \bibnamefont
  {Cirac}},\ }\bibfield  {title} {\enquote {\bibinfo {title} {{Density Matrix
  Renormalization Group and Periodic Boundary Conditions: A Quantum Information
  Perspective}},}\ }\href {\doibase 10.1103/PhysRevLett.93.227205} {\bibfield
  {journal} {\bibinfo  {journal} {Phys. Rev. Lett.}\ }\textbf {\bibinfo
  {volume} {93}},\ \bibinfo {pages} {227205} (\bibinfo {year}
  {2004}{\natexlab{b}})}\BibitemShut {NoStop}%
\bibitem [{\citenamefont {Gross}\ \emph {et~al.}(2007)\citenamefont {Gross},
  \citenamefont {Eisert}, \citenamefont {Schuch},\ and\ \citenamefont
  {Perez-Garcia}}]{GESP07MPSQC}%
  \BibitemOpen
  \bibfield  {author} {\bibinfo {author} {\bibfnamefont {D.}~\bibnamefont
  {Gross}}, \bibinfo {author} {\bibfnamefont {J.}~\bibnamefont {Eisert}},
  \bibinfo {author} {\bibfnamefont {N.}~\bibnamefont {Schuch}}, \ and\ \bibinfo
  {author} {\bibfnamefont {D.}~\bibnamefont {Perez-Garcia}},\ }\bibfield
  {title} {\enquote {\bibinfo {title} {Measurement-based quantum computation
  beyond the one-way model},}\ }\href {\doibase 10.1103/PhysRevA.76.052315}
  {\bibfield  {journal} {\bibinfo  {journal} {Phys. Rev. A}\ }\textbf {\bibinfo
  {volume} {76}},\ \bibinfo {pages} {052315} (\bibinfo {year}
  {2007})}\BibitemShut {NoStop}%
\bibitem [{\citenamefont {Verstraete}\ \emph {et~al.}(2009)\citenamefont
  {Verstraete}, \citenamefont {Wolf},\ and\ \citenamefont {Cirac}}]{VWC09QCP}%
  \BibitemOpen
  \bibfield  {author} {\bibinfo {author} {\bibfnamefont {Frank}\ \bibnamefont
  {Verstraete}}, \bibinfo {author} {\bibfnamefont {Michael~M}\ \bibnamefont
  {Wolf}}, \ and\ \bibinfo {author} {\bibfnamefont {J~Ignacio}\ \bibnamefont
  {Cirac}},\ }\bibfield  {title} {\enquote {\bibinfo {title} {Quantum
  computation and quantum-state engineering driven by dissipation},}\ }\href
  {\doibase 10.1038/nphys1342} {\bibfield  {journal} {\bibinfo  {journal} {Nat.
  phys.}\ }\textbf {\bibinfo {volume} {5}},\ \bibinfo {pages} {633} (\bibinfo
  {year} {2009})}\BibitemShut {NoStop}%
\bibitem [{\citenamefont {Dang}\ \emph {et~al.}(2019)\citenamefont {Dang},
  \citenamefont {Hill},\ and\ \citenamefont {Hollenberg}}]{DHH19MPSshor}%
  \BibitemOpen
  \bibfield  {author} {\bibinfo {author} {\bibfnamefont {Aidan}\ \bibnamefont
  {Dang}}, \bibinfo {author} {\bibfnamefont {Charles~D}\ \bibnamefont {Hill}},
  \ and\ \bibinfo {author} {\bibfnamefont {Lloyd~CL}\ \bibnamefont
  {Hollenberg}},\ }\bibfield  {title} {\enquote {\bibinfo {title} {Optimising
  matrix product state simulations of shor’s algorithm},}\ }\href {\doibase
  10.22331/q-2019-01-25-116} {\bibfield  {journal} {\bibinfo  {journal}
  {Quantum}\ }\textbf {\bibinfo {volume} {3}},\ \bibinfo {pages} {116}
  (\bibinfo {year} {2019})}\BibitemShut {NoStop}%
\bibitem [{\citenamefont {Bhatia}\ and\ \citenamefont
  {Saggi}(2018)}]{BS18MPSQC}%
  \BibitemOpen
  \bibfield  {author} {\bibinfo {author} {\bibfnamefont {Amandeep~Singh}\
  \bibnamefont {Bhatia}}\ and\ \bibinfo {author} {\bibfnamefont {Mandeep~Kaur}\
  \bibnamefont {Saggi}},\ }\bibfield  {title} {\enquote {\bibinfo {title}
  {Simulation of matrix product state on a quantum computer},}\ }\href@noop {}
  {\  (\bibinfo {year} {2018})},\ \Eprint
  {http://arxiv.org/abs/arXiv:1811.09833} {arXiv:1811.09833} \BibitemShut
  {NoStop}%
\bibitem [{\citenamefont {Affleck}\ \emph {et~al.}(1987)\citenamefont
  {Affleck}, \citenamefont {Kennedy}, \citenamefont {Lieb},\ and\ \citenamefont
  {Tasaki}}]{AKLT87AKLTState}%
  \BibitemOpen
  \bibfield  {author} {\bibinfo {author} {\bibfnamefont {Ian}\ \bibnamefont
  {Affleck}}, \bibinfo {author} {\bibfnamefont {Tom}\ \bibnamefont {Kennedy}},
  \bibinfo {author} {\bibfnamefont {Elliott~H.}\ \bibnamefont {Lieb}}, \ and\
  \bibinfo {author} {\bibfnamefont {Hal}\ \bibnamefont {Tasaki}},\ }\bibfield
  {title} {\enquote {\bibinfo {title} {Rigorous results on valence-bond ground
  states in antiferromagnets},}\ }\href {\doibase 10.1103/PhysRevLett.59.799}
  {\bibfield  {journal} {\bibinfo  {journal} {Phys. Rev. Lett.}\ }\textbf
  {\bibinfo {volume} {59}},\ \bibinfo {pages} {799--802} (\bibinfo {year}
  {1987})}\BibitemShut {NoStop}%
\bibitem [{\citenamefont {P\'{e}rez-Garc\'ia}\ \emph
  {et~al.}(2007)\citenamefont {P\'{e}rez-Garc\'ia}, \citenamefont {Verstraete},
  \citenamefont {Wolf},\ and\ \citenamefont {Cirac}}]{PVWC07MPSRev}%
  \BibitemOpen
  \bibfield  {author} {\bibinfo {author} {\bibfnamefont {David}\ \bibnamefont
  {P\'{e}rez-Garc\'ia}}, \bibinfo {author} {\bibfnamefont {Frank}\ \bibnamefont
  {Verstraete}}, \bibinfo {author} {\bibfnamefont {Michael~M.}\ \bibnamefont
  {Wolf}}, \ and\ \bibinfo {author} {\bibfnamefont {J.~Ignacio}\ \bibnamefont
  {Cirac}},\ }\bibfield  {title} {\enquote {\bibinfo {title} {{Matrix Product
  State Representations}},}\ }\href@noop {} {\bibfield  {journal} {\bibinfo
  {journal} {Quantum Inf. Comput.}\ }\textbf {\bibinfo {volume} {7}},\ \bibinfo
  {pages} {401--430} (\bibinfo {year} {2007})}\BibitemShut {NoStop}%
\bibitem [{\citenamefont {Verstraete}\ \emph
  {et~al.}(2004{\natexlab{c}})\citenamefont {Verstraete}, \citenamefont
  {Mart\'{\i}n-Delgado},\ and\ \citenamefont {Cirac}}]{PhysRevLett.92.087201}%
  \BibitemOpen
  \bibfield  {author} {\bibinfo {author} {\bibfnamefont {F.}~\bibnamefont
  {Verstraete}}, \bibinfo {author} {\bibfnamefont {M.~A.}\ \bibnamefont
  {Mart\'{\i}n-Delgado}}, \ and\ \bibinfo {author} {\bibfnamefont {J.~I.}\
  \bibnamefont {Cirac}},\ }\bibfield  {title} {\enquote {\bibinfo {title}
  {Diverging entanglement length in gapped quantum spin systems},}\ }\href
  {\doibase 10.1103/PhysRevLett.92.087201} {\bibfield  {journal} {\bibinfo
  {journal} {Phys. Rev. Lett.}\ }\textbf {\bibinfo {volume} {92}},\ \bibinfo
  {pages} {087201} (\bibinfo {year} {2004}{\natexlab{c}})}\BibitemShut
  {NoStop}%
\bibitem [{\citenamefont {Gross}\ and\ \citenamefont
  {Eisert}(2007)}]{GE07QCPEPS}%
  \BibitemOpen
  \bibfield  {author} {\bibinfo {author} {\bibfnamefont {D.}~\bibnamefont
  {Gross}}\ and\ \bibinfo {author} {\bibfnamefont {J.}~\bibnamefont {Eisert}},\
  }\bibfield  {title} {\enquote {\bibinfo {title} {Novel schemes for
  measurement-based quantum computation},}\ }\href {\doibase
  10.1103/PhysRevLett.98.220503} {\bibfield  {journal} {\bibinfo  {journal}
  {Phys. Rev. Lett.}\ }\textbf {\bibinfo {volume} {98}},\ \bibinfo {pages}
  {220503} (\bibinfo {year} {2007})}\BibitemShut {NoStop}%
\bibitem [{\citenamefont {Wei}\ \emph {et~al.}(2011)\citenamefont {Wei},
  \citenamefont {Affleck},\ and\ \citenamefont {Raussendorf}}]{WAR11AKLTQC}%
  \BibitemOpen
  \bibfield  {author} {\bibinfo {author} {\bibfnamefont {Tzu-Chieh}\
  \bibnamefont {Wei}}, \bibinfo {author} {\bibfnamefont {Ian}\ \bibnamefont
  {Affleck}}, \ and\ \bibinfo {author} {\bibfnamefont {Robert}\ \bibnamefont
  {Raussendorf}},\ }\bibfield  {title} {\enquote {\bibinfo {title}
  {Affleck-kennedy-lieb-tasaki state on a honeycomb lattice is a universal
  quantum computational resource},}\ }\href {\doibase
  10.1103/PhysRevLett.106.070501} {\bibfield  {journal} {\bibinfo  {journal}
  {Phys. Rev. Lett.}\ }\textbf {\bibinfo {volume} {106}},\ \bibinfo {pages}
  {070501} (\bibinfo {year} {2011})}\BibitemShut {NoStop}%
\bibitem [{\citenamefont {Else}\ \emph {et~al.}(2012)\citenamefont {Else},
  \citenamefont {Schwarz}, \citenamefont {Bartlett},\ and\ \citenamefont
  {Doherty}}]{ESBD12AKLTQC}%
  \BibitemOpen
  \bibfield  {author} {\bibinfo {author} {\bibfnamefont {Dominic~V.}\
  \bibnamefont {Else}}, \bibinfo {author} {\bibfnamefont {Ilai}\ \bibnamefont
  {Schwarz}}, \bibinfo {author} {\bibfnamefont {Stephen~D.}\ \bibnamefont
  {Bartlett}}, \ and\ \bibinfo {author} {\bibfnamefont {Andrew~C.}\
  \bibnamefont {Doherty}},\ }\bibfield  {title} {\enquote {\bibinfo {title}
  {Symmetry-protected phases for measurement-based quantum computation},}\
  }\href {\doibase 10.1103/PhysRevLett.108.240505} {\bibfield  {journal}
  {\bibinfo  {journal} {Phys. Rev. Lett.}\ }\textbf {\bibinfo {volume} {108}},\
  \bibinfo {pages} {240505} (\bibinfo {year} {2012})}\BibitemShut {NoStop}%
\bibitem [{\citenamefont {Song}\ \emph {et~al.}(2019)\citenamefont {Song},
  \citenamefont {Xu}, \citenamefont {Li}, \citenamefont {Zhang}, \citenamefont
  {Zhang}, \citenamefont {Liu}, \citenamefont {Guo}, \citenamefont {Wang},
  \citenamefont {Ren}, \citenamefont {Hao} \emph {et~al.}}]{SXLZ+19GHZ20}%
  \BibitemOpen
  \bibfield  {author} {\bibinfo {author} {\bibfnamefont {Chao}\ \bibnamefont
  {Song}}, \bibinfo {author} {\bibfnamefont {Kai}\ \bibnamefont {Xu}}, \bibinfo
  {author} {\bibfnamefont {Hekang}\ \bibnamefont {Li}}, \bibinfo {author}
  {\bibfnamefont {Yu-Ran}\ \bibnamefont {Zhang}}, \bibinfo {author}
  {\bibfnamefont {Xu}~\bibnamefont {Zhang}}, \bibinfo {author} {\bibfnamefont
  {Wuxin}\ \bibnamefont {Liu}}, \bibinfo {author} {\bibfnamefont {Qiujiang}\
  \bibnamefont {Guo}}, \bibinfo {author} {\bibfnamefont {Zhen}\ \bibnamefont
  {Wang}}, \bibinfo {author} {\bibfnamefont {Wenhui}\ \bibnamefont {Ren}},
  \bibinfo {author} {\bibfnamefont {Jie}\ \bibnamefont {Hao}},  \emph
  {et~al.},\ }\bibfield  {title} {\enquote {\bibinfo {title} {Generation of
  multicomponent atomic schr{\"o}dinger cat states of up to 20 qubits},}\
  }\href {\doibase 10.1126/science.aay0600} {\bibfield  {journal} {\bibinfo
  {journal} {Science}\ }\textbf {\bibinfo {volume} {365}},\ \bibinfo {pages}
  {574--577} (\bibinfo {year} {2019})}\BibitemShut {NoStop}%
\bibitem [{\citenamefont {Omran}\ \emph {et~al.}(2019)\citenamefont {Omran},
  \citenamefont {Levine}, \citenamefont {Keesling}, \citenamefont {Semeghini},
  \citenamefont {Wang}, \citenamefont {Ebadi}, \citenamefont {Bernien},
  \citenamefont {Zibrov}, \citenamefont {Pichler}, \citenamefont {Choi},
  \citenamefont {Cui}, \citenamefont {Rossignolo}, \citenamefont {Rembold},
  \citenamefont {Montangero}, \citenamefont {Calarco}, \citenamefont {Endres},
  \citenamefont {Greiner}, \citenamefont {Vuleti{\'c}},\ and\ \citenamefont
  {Lukin}}]{OLKS+19CatExp}%
  \BibitemOpen
  \bibfield  {author} {\bibinfo {author} {\bibfnamefont {A.}~\bibnamefont
  {Omran}}, \bibinfo {author} {\bibfnamefont {H.}~\bibnamefont {Levine}},
  \bibinfo {author} {\bibfnamefont {A.}~\bibnamefont {Keesling}}, \bibinfo
  {author} {\bibfnamefont {G.}~\bibnamefont {Semeghini}}, \bibinfo {author}
  {\bibfnamefont {T.~T.}\ \bibnamefont {Wang}}, \bibinfo {author}
  {\bibfnamefont {S.}~\bibnamefont {Ebadi}}, \bibinfo {author} {\bibfnamefont
  {H.}~\bibnamefont {Bernien}}, \bibinfo {author} {\bibfnamefont {A.~S.}\
  \bibnamefont {Zibrov}}, \bibinfo {author} {\bibfnamefont {H.}~\bibnamefont
  {Pichler}}, \bibinfo {author} {\bibfnamefont {S.}~\bibnamefont {Choi}},
  \bibinfo {author} {\bibfnamefont {J.}~\bibnamefont {Cui}}, \bibinfo {author}
  {\bibfnamefont {M.}~\bibnamefont {Rossignolo}}, \bibinfo {author}
  {\bibfnamefont {P.}~\bibnamefont {Rembold}}, \bibinfo {author} {\bibfnamefont
  {S.}~\bibnamefont {Montangero}}, \bibinfo {author} {\bibfnamefont
  {T.}~\bibnamefont {Calarco}}, \bibinfo {author} {\bibfnamefont
  {M.}~\bibnamefont {Endres}}, \bibinfo {author} {\bibfnamefont
  {M.}~\bibnamefont {Greiner}}, \bibinfo {author} {\bibfnamefont
  {V.}~\bibnamefont {Vuleti{\'c}}}, \ and\ \bibinfo {author} {\bibfnamefont
  {M.~D.}\ \bibnamefont {Lukin}},\ }\bibfield  {title} {\enquote {\bibinfo
  {title} {Generation and manipulation of schr{\"o}dinger cat states in rydberg
  atom arrays},}\ }\href {\doibase 10.1126/science.aax9743} {\bibfield
  {journal} {\bibinfo  {journal} {Science}\ }\textbf {\bibinfo {volume}
  {365}},\ \bibinfo {pages} {570--574} (\bibinfo {year} {2019})}\BibitemShut
  {NoStop}%
\bibitem [{\citenamefont {Sch\"on}\ \emph {et~al.}(2005)\citenamefont
  {Sch\"on}, \citenamefont {Solano}, \citenamefont {Verstraete}, \citenamefont
  {Cirac},\ and\ \citenamefont {Wolf}}]{PhysRevLett.95.110503}%
  \BibitemOpen
  \bibfield  {author} {\bibinfo {author} {\bibfnamefont {C.}~\bibnamefont
  {Sch\"on}}, \bibinfo {author} {\bibfnamefont {E.}~\bibnamefont {Solano}},
  \bibinfo {author} {\bibfnamefont {F.}~\bibnamefont {Verstraete}}, \bibinfo
  {author} {\bibfnamefont {J.~I.}\ \bibnamefont {Cirac}}, \ and\ \bibinfo
  {author} {\bibfnamefont {M.~M.}\ \bibnamefont {Wolf}},\ }\bibfield  {title}
  {\enquote {\bibinfo {title} {Sequential generation of entangled multiqubit
  states},}\ }\href {\doibase 10.1103/PhysRevLett.95.110503} {\bibfield
  {journal} {\bibinfo  {journal} {Phys. Rev. Lett.}\ }\textbf {\bibinfo
  {volume} {95}},\ \bibinfo {pages} {110503} (\bibinfo {year}
  {2005})}\BibitemShut {NoStop}%
\bibitem [{\citenamefont {Cramer}\ \emph {et~al.}(2010)\citenamefont {Cramer},
  \citenamefont {Plenio}, \citenamefont {Flammia}, \citenamefont {Somma},
  \citenamefont {Gross}, \citenamefont {Bartlett}, \citenamefont
  {Landon-Cardinal}, \citenamefont {Poulin},\ and\ \citenamefont
  {Liu}}]{cramer2010efficient}%
  \BibitemOpen
  \bibfield  {author} {\bibinfo {author} {\bibfnamefont {Marcus}\ \bibnamefont
  {Cramer}}, \bibinfo {author} {\bibfnamefont {Martin~B}\ \bibnamefont
  {Plenio}}, \bibinfo {author} {\bibfnamefont {Steven~T}\ \bibnamefont
  {Flammia}}, \bibinfo {author} {\bibfnamefont {Rolando}\ \bibnamefont
  {Somma}}, \bibinfo {author} {\bibfnamefont {David}\ \bibnamefont {Gross}},
  \bibinfo {author} {\bibfnamefont {Stephen~D}\ \bibnamefont {Bartlett}},
  \bibinfo {author} {\bibfnamefont {Olivier}\ \bibnamefont {Landon-Cardinal}},
  \bibinfo {author} {\bibfnamefont {David}\ \bibnamefont {Poulin}}, \ and\
  \bibinfo {author} {\bibfnamefont {Yi-Kai}\ \bibnamefont {Liu}},\ }\bibfield
  {title} {\enquote {\bibinfo {title} {Efficient quantum state tomography},}\
  }\href {\doibase 10.1038/ncomms1147} {\bibfield  {journal} {\bibinfo
  {journal} {Nat. Comm.}\ }\textbf {\bibinfo {volume} {1}},\ \bibinfo {pages}
  {149} (\bibinfo {year} {2010})}\BibitemShut {NoStop}%
\bibitem [{\citenamefont {Barenco}\ \emph {et~al.}(1995)\citenamefont
  {Barenco}, \citenamefont {Bennett}, \citenamefont {Cleve}, \citenamefont
  {DiVincenzo}, \citenamefont {Margolus}, \citenamefont {Shor}, \citenamefont
  {Sleator}, \citenamefont {Smolin},\ and\ \citenamefont
  {Weinfurter}}]{BBCD+95universalGate}%
  \BibitemOpen
  \bibfield  {author} {\bibinfo {author} {\bibfnamefont {Adriano}\ \bibnamefont
  {Barenco}}, \bibinfo {author} {\bibfnamefont {Charles~H.}\ \bibnamefont
  {Bennett}}, \bibinfo {author} {\bibfnamefont {Richard}\ \bibnamefont
  {Cleve}}, \bibinfo {author} {\bibfnamefont {David~P.}\ \bibnamefont
  {DiVincenzo}}, \bibinfo {author} {\bibfnamefont {Norman}\ \bibnamefont
  {Margolus}}, \bibinfo {author} {\bibfnamefont {Peter}\ \bibnamefont {Shor}},
  \bibinfo {author} {\bibfnamefont {Tycho}\ \bibnamefont {Sleator}}, \bibinfo
  {author} {\bibfnamefont {John~A.}\ \bibnamefont {Smolin}}, \ and\ \bibinfo
  {author} {\bibfnamefont {Harald}\ \bibnamefont {Weinfurter}},\ }\bibfield
  {title} {\enquote {\bibinfo {title} {Elementary gates for quantum
  computation},}\ }\href {\doibase 10.1103/PhysRevA.52.3457} {\bibfield
  {journal} {\bibinfo  {journal} {Phys. Rev. A}\ }\textbf {\bibinfo {volume}
  {52}},\ \bibinfo {pages} {3457--3467} (\bibinfo {year} {1995})}\BibitemShut
  {NoStop}%
\bibitem [{\citenamefont {Chong}\ \emph {et~al.}(2017)\citenamefont {Chong},
  \citenamefont {Franklin},\ and\ \citenamefont {Martonosi}}]{CFM17Qcompile}%
  \BibitemOpen
  \bibfield  {author} {\bibinfo {author} {\bibfnamefont {Frederic~T}\
  \bibnamefont {Chong}}, \bibinfo {author} {\bibfnamefont {Diana}\ \bibnamefont
  {Franklin}}, \ and\ \bibinfo {author} {\bibfnamefont {Margaret}\ \bibnamefont
  {Martonosi}},\ }\bibfield  {title} {\enquote {\bibinfo {title} {Programming
  languages and compiler design for realistic quantum hardware},}\ }\href
  {\doibase 10.1038/nature23459} {\bibfield  {journal} {\bibinfo  {journal}
  {Nature}\ }\textbf {\bibinfo {volume} {549}},\ \bibinfo {pages} {180}
  (\bibinfo {year} {2017})}\BibitemShut {NoStop}%
\bibitem [{\citenamefont {Liu}\ \emph {et~al.}(2019)\citenamefont {Liu},
  \citenamefont {Zhang}, \citenamefont {Wan},\ and\ \citenamefont
  {Wang}}]{LZWW19QMPS}%
  \BibitemOpen
  \bibfield  {author} {\bibinfo {author} {\bibfnamefont {Jin-Guo}\ \bibnamefont
  {Liu}}, \bibinfo {author} {\bibfnamefont {Yi-Hong}\ \bibnamefont {Zhang}},
  \bibinfo {author} {\bibfnamefont {Yuan}\ \bibnamefont {Wan}}, \ and\ \bibinfo
  {author} {\bibfnamefont {Lei}\ \bibnamefont {Wang}},\ }\bibfield  {title}
  {\enquote {\bibinfo {title} {Variational quantum eigensolver with fewer
  qubits},}\ }\href {\doibase 10.1103/PhysRevResearch.1.023025} {\bibfield
  {journal} {\bibinfo  {journal} {Phys. Rev. Research}\ }\textbf {\bibinfo
  {volume} {1}},\ \bibinfo {pages} {023025} (\bibinfo {year}
  {2019})}\BibitemShut {NoStop}%
\bibitem [{\citenamefont {Mottonen}\ and\ \citenamefont
  {Vartiainen}(2006)}]{MV06Qgates}%
  \BibitemOpen
  \bibfield  {author} {\bibinfo {author} {\bibfnamefont {M.}~\bibnamefont
  {Mottonen}}\ and\ \bibinfo {author} {\bibfnamefont {J.~J.}\ \bibnamefont
  {Vartiainen}},\ }\enquote {\bibinfo {title} {Decompositions of general
  quantum gates},}\ in\ \href@noop {} {\emph {\bibinfo {booktitle} {Trends in
  Quantum Computing Research}}},\ \bibinfo {editor} {edited by\ \bibinfo
  {editor} {\bibfnamefont {Susan}\ \bibnamefont {Shannon}}}\ (\bibinfo
  {publisher} {NOVA Publishers},\ \bibinfo {address} {New York},\ \bibinfo
  {year} {2006})\ p.\ \bibinfo {pages} {Chapter 7}\BibitemShut {NoStop}%
\bibitem [{\citenamefont {Cirac}\ \emph {et~al.}(2017)\citenamefont {Cirac},
  \citenamefont {Perez-Garcia}, \citenamefont {Schuch},\ and\ \citenamefont
  {Verstraete}}]{CPSV17UMPO}%
  \BibitemOpen
  \bibfield  {author} {\bibinfo {author} {\bibfnamefont {J~Ignacio}\
  \bibnamefont {Cirac}}, \bibinfo {author} {\bibfnamefont {David}\ \bibnamefont
  {Perez-Garcia}}, \bibinfo {author} {\bibfnamefont {Norbert}\ \bibnamefont
  {Schuch}}, \ and\ \bibinfo {author} {\bibfnamefont {Frank}\ \bibnamefont
  {Verstraete}},\ }\bibfield  {title} {\enquote {\bibinfo {title} {Matrix
  product unitaries: structure, symmetries, and topological invariants},}\
  }\href {\doibase 10.1088/1742-5468/aa7e55} {\bibfield  {journal} {\bibinfo
  {journal} {J. Stat. Mech.}\ }\textbf {\bibinfo {volume} {2017}},\ \bibinfo
  {pages} {083105} (\bibinfo {year} {2017})}\BibitemShut {NoStop}%
\bibitem [{\citenamefont {\ifmmode \mbox{\c{S}}\else
  \c{S}\fi{}ahino\ifmmode~\breve{g}\else \u{g}\fi{}lu}\ \emph
  {et~al.}(2018)\citenamefont {\ifmmode \mbox{\c{S}}\else
  \c{S}\fi{}ahino\ifmmode~\breve{g}\else \u{g}\fi{}lu}, \citenamefont {Shukla},
  \citenamefont {Bi},\ and\ \citenamefont {Chen}}]{SBS+18UMPO}%
  \BibitemOpen
  \bibfield  {author} {\bibinfo {author} {\bibfnamefont {M.~Burak}\
  \bibnamefont {\ifmmode \mbox{\c{S}}\else
  \c{S}\fi{}ahino\ifmmode~\breve{g}\else \u{g}\fi{}lu}}, \bibinfo {author}
  {\bibfnamefont {Sujeet~K.}\ \bibnamefont {Shukla}}, \bibinfo {author}
  {\bibfnamefont {Feng}\ \bibnamefont {Bi}}, \ and\ \bibinfo {author}
  {\bibfnamefont {Xie}\ \bibnamefont {Chen}},\ }\bibfield  {title} {\enquote
  {\bibinfo {title} {Matrix product representation of locality preserving
  unitaries},}\ }\href {\doibase 10.1103/PhysRevB.98.245122} {\bibfield
  {journal} {\bibinfo  {journal} {Phys. Rev. B}\ }\textbf {\bibinfo {volume}
  {98}},\ \bibinfo {pages} {245122} (\bibinfo {year} {2018})}\BibitemShut
  {NoStop}%
\bibitem [{\citenamefont {Vidal}\ \emph {et~al.}(2003)\citenamefont {Vidal},
  \citenamefont {Latorre}, \citenamefont {Rico},\ and\ \citenamefont
  {Kitaev}}]{VLRK03CritEnt}%
  \BibitemOpen
  \bibfield  {author} {\bibinfo {author} {\bibfnamefont {Guifr\'{e}}\
  \bibnamefont {Vidal}}, \bibinfo {author} {\bibfnamefont {Jos\'{e}~I.}\
  \bibnamefont {Latorre}}, \bibinfo {author} {\bibfnamefont {Enrique}\
  \bibnamefont {Rico}}, \ and\ \bibinfo {author} {\bibfnamefont {Alexei}\
  \bibnamefont {Kitaev}},\ }\bibfield  {title} {\enquote {\bibinfo {title}
  {{Entanglement in Quantum Critical Phenomena}},}\ }\href {\doibase
  10.1103/PhysRevLett.90.227902} {\bibfield  {journal} {\bibinfo  {journal}
  {Phys. Rev. Lett.}\ }\textbf {\bibinfo {volume} {90}},\ \bibinfo {pages}
  {227902} (\bibinfo {year} {2003})}\BibitemShut {NoStop}%
\bibitem [{\citenamefont {Tagliacozzo}\ \emph {et~al.}(2008)\citenamefont
  {Tagliacozzo}, \citenamefont {de~Oliveira}, \citenamefont {Iblisdir},\ and\
  \citenamefont {Latorre}}]{TOIL08EntScaling}%
  \BibitemOpen
  \bibfield  {author} {\bibinfo {author} {\bibfnamefont {Luca}\ \bibnamefont
  {Tagliacozzo}}, \bibinfo {author} {\bibfnamefont {Thiago~R.}\ \bibnamefont
  {de~Oliveira}}, \bibinfo {author} {\bibfnamefont {Sofyan}\ \bibnamefont
  {Iblisdir}}, \ and\ \bibinfo {author} {\bibfnamefont {Jos\'{e}~I.}\
  \bibnamefont {Latorre}},\ }\bibfield  {title} {\enquote {\bibinfo {title}
  {{Scaling of entanglement support for matrix product states}},}\ }\href
  {\doibase 10.1103/PhysRevB.78.024410} {\bibfield  {journal} {\bibinfo
  {journal} {Phys. Rev. B}\ }\textbf {\bibinfo {volume} {78}},\ \bibinfo
  {pages} {024410} (\bibinfo {year} {2008})}\BibitemShut {NoStop}%
\bibitem [{\citenamefont {Schollw\"{o}ck}(2011)}]{S11DMRGRev}%
  \BibitemOpen
  \bibfield  {author} {\bibinfo {author} {\bibfnamefont {Ulrich}\ \bibnamefont
  {Schollw\"{o}ck}},\ }\bibfield  {title} {\enquote {\bibinfo {title} {{The
  density-matrix renormalization group in the age of matrix product states}},}\
  }\href {\doibase 10.1016/j.aop.2010.09.012} {\bibfield  {journal} {\bibinfo
  {journal} {Ann. Phys.}\ }\textbf {\bibinfo {volume} {326}},\ \bibinfo {pages}
  {96--192} (\bibinfo {year} {2011})}\BibitemShut {NoStop}%
\bibitem [{not()}]{note1}%
  \BibitemOpen
  \href@noop {} {}\bibinfo {note} {See the Appendix for the proof of $\hat{U}$
  being unitary.}\BibitemShut {Stop}%
\bibitem [{\citenamefont {Wei}\ and\ \citenamefont {Goldbart}(2003)}]{WG03GE}%
  \BibitemOpen
  \bibfield  {author} {\bibinfo {author} {\bibfnamefont {Tzu-Chieh}\
  \bibnamefont {Wei}}\ and\ \bibinfo {author} {\bibfnamefont {Paul~M.}\
  \bibnamefont {Goldbart}},\ }\bibfield  {title} {\enquote {\bibinfo {title}
  {Geometric measure of entanglement and applications to bipartite and
  multipartite quantum states},}\ }\href {\doibase 10.1103/PhysRevA.68.042307}
  {\bibfield  {journal} {\bibinfo  {journal} {Phys. Rev. A}\ }\textbf {\bibinfo
  {volume} {68}},\ \bibinfo {pages} {042307} (\bibinfo {year}
  {2003})}\BibitemShut {NoStop}%
\end{thebibliography}

\end{document}